\RequirePackage{ifpdf}
\documentclass[12pt,letterpaper]{JHEP3}
\usepackage{cite}
\usepackage{epsfig}
\usepackage{graphicx}
\usepackage{subfigure}

\usepackage{color}
\newcommand{\Beq}{\begin{eqnarray}}
\newcommand{\Eeq}{\end{eqnarray}}
\newcommand{\tint}{t_{\rm{int}}}

\newcommand{\Aint}{\mathcal{A}_{\rm{int}}}

\newcommand{\bx}{{\bf{x}}}
\newcommand{\mG}{{\mathcal{G}}}

\newcommand{\eqn}[1]{equation (\ref{#1})}
\newcommand{\nn}{\nonumber \\}

\let\normalcolor\relax 
\title{A scattering theory of ultrarelativistic solitons}

\author{Mustafa A. Amin\footnote{mustafa.a.amin@gmail.com}, Eugene A. Lim\footnote{eugene.a.lim@gmail.com} and 
I-Sheng Yang\footnote{isheng.yang@gmail.com} \\
${}^{*}$ Kavli Institute for Cosmology and Institute of Astronomy, Madingley Rd, Cambridge CB3 0HA, United Kingdom\\
${}^{\dagger}$ Theoretical Particle Physics and Cosmology Group, Physics Department,
King's College London, Strand, London WC2R 2LS, United Kingdom\\
${}^{\ddagger}$IOP and GRAPPA, Universiteit van Amsterdam, Science Park 904, 1090 GL Amsterdam, Netherlands
}

\abstract
{
We construct a perturbative framework for understanding the collision of solitons (more precisely, solitary waves) in relativistic scalar field theories. Our perturbative framework is based on the suppression of the space-time interaction area proportional to $1/(\gamma v)$, where $v$ is the relative velocity of an incoming solitary wave and $\gamma=1/\sqrt{1-v^2}\gg 1$. We calculate the leading order results for collisions of (1+1) dimensional kinks in periodic potentials, and provide explicit, closed form expressions for the phase shift and the velocity change after the collisions. We find excellent agreement between our results and detailed numerical simulations. Crucially, our perturbation series is controlled by a kinematic parameter, and hence not restricted to small deviations around integrable cases such as the Sine-Gordon model. %We comment on the possible applications and directions for further studies.
} 
\begin{document}
%~~~~~~~~~~~~~~~~ Introduction ~~~~~~~~~~~~~~~~~~~~~~~~~~~~~~~~
%%%%%%%%%%%%%%%%%%%%%%%%%%%%%%%%%%%%%%%%%%%%%%%%
\section{Introduction}
In 1834, J. S. Russell discovered solitary waves \cite{Russell:1844}: spatially localized configurations of fields that propagate without any distortion.  More than 150 years after their discovery, solitary waves are still actively studied for their often counter-intuitive, yet elegant properties \cite{ScoChu73,SWRev80,Raj82}. They have found applications in disparate fields such as atomic physics\cite{IshHid84}, superconductivity\cite{Fio05}, field theory\cite{Ria02}, biology\cite{Yak05}, condensed matter physics\cite{GogNer99,Log04}, nonlinear optics\cite{BulCau80}, quantum chromodynamics\cite{Bla07}, cosmology \cite{Kle11, AguJoh09, Amin:2011hj} and neuroscience\cite{VilLud10}. Our paper is an attempt to further the study of interactions of these fascinating objects. 

Solitary waves have a particular subset called solitons\cite{ScoChu73} which possess a rather intriguing property. Solitons are not only stable on their own, collisions between solitons leave them completely unchanged apart from a phase shift. A tremendous amount of work exists on understanding solitons and their interactions\cite{Gar67,Lax68,McLAlw73,McLSco78,Mal85a,Mal85b,KivMal89}. However, examples of systems with true solitons are limited (for example, see \cite{ScoChu73}). Such systems are often integrable and for such integrable systems powerful techniques exist to write down their multi-soliton solutions analytically \cite{Gar67,Lax68}. However, it is difficult to know a-priori whether the solitary waves in a theory are also solitons and whether such systems admit analytic multi-soliton solutions.  

We would like to predict the results of general solitary wave collisions. Such collisions include, but are not limited to the elastic case, characteristic of true solitons in integrable systems. We focus on solitary waves in relativistic scalar field theories with canonical kinetic terms and potentials with effectively single minima (including periodic potentials with multiple minima). Such theories and their solitary wave solutions appear naturally in cosmology and high energy physics\cite{Wei12,Mantonbook}.\footnote{Some condensed matter systems near critical points also admit a similar structure\cite{He3}. One simply replace the speed of light by the Fermi velocity\cite{YanTye11,MasXia12}.} We are also motivated by the fact that in certain relativistic scalar field theories, the solitary waves effectively pass through each other when colliding at ultrarelativistic velocities \cite{She87, Axenides:1999hs, Hindmarsh:2007jb, EasGib09,GibLam10,DesGib12}.

In this paper, {\it we provide a general perturbative framework to study ultrarelativistic solitary wave collisions}. Assuming that the colliding solitary waves effectively pass through each other at zeroth order \footnote{In the case of single minimum or periodic potentials, effectively passing through each other is captured by a linear superposition of the two solitary wave solutions. In multiple minima potentials which are not periodic, this is not true and further modification of the solitary wave profiles has to be taken into account. As a result, we restrict ourselves to single minimum or periodic potentials in this paper.}, we provide a perturbative framework to calculate the corrections to this `free passage' behavior. We then show that the corrections are indeed small and present an order by order prescription to calculate the full result.

The essential ingredient of our perturbative framework is as follows. The colliding solitary waves interact significantly only when they overlap. In the rest frame of one solitary wave where the other approaches with velocity $v\rightarrow 1$, the space-time area of such overlap $\Aint\propto (\gamma v)^{-1}$ where $\gamma =(1-v^2)^{-1/2}\gg1$. We use this property to provide a perturbative framework to calculate the effects of collisions, with $(\gamma v)^{-1}$ as the small expansion parameter. 

In Sec.\ref{sec-GPertExp} we discuss the conditions that the solitary waves must satisfy for our framework to be applicable and state some simplifying assumptions. Despite these simplifications, our framework should be applicable to many well-known objects like oscillons\cite{CopGle95}, Q-balls\cite{Col85}, and domain walls. We then present the general framework of our perturbation theory to calculate the effects of the collision, paying particular attention to how the $(\gamma v)^{-1}$ expansion emerges. Since we are interested in the effect of the collisions, not necessarily the subsequent evolution of the perturbations, we always evaluate the perturbations soon after the collision.  The long time behavior of the perturbations is discussed in Appendix \ref{sec-LongTime}. For simplicity, we only consider linearized perturbations in the main body of the text.  Nonlinear effects are discussed in Appendix \ref{sec-nonlinear}. 

After this general discussion, in Sec. \ref{sec-kinks} and Sec. \ref{sec-pert} we focus on a simple example in (1+1) dimensions---kinks\footnote{A stationary kink is a static solution that interpolates between adjacent minima of the potential. In higher dimensions they would be domain walls.} in a single scalar field theory with periodic potentials: $V(\phi)=V(\phi+\Delta\phi)$ for all $\phi$.  We explicitly calculate the effects of the kink-kink collision at leading order in our expansion parameter $(\gamma v)^{-1}$.  These leading order results already reveal a few important and surprising facts. In the rest frame of the stationary kink which collided with an incoming kink, we find the following:
\begin{itemize}
\item The velocity change of the stationary kink is zero (therefore its velocity remains zero).
\begin{eqnarray}
\Delta v = 0+\mathcal{O}[(\gamma v)^{-2}]~.
\end{eqnarray}
\item The stationary kink acquires a phase shift (spatial translation)\footnote{Note that the spatial translation is only meaningful in the absence of the velocity change.} given by
\begin{eqnarray}
\Delta x
&=&\frac{1}{2(\gamma v) M}\int_{0}^{\Delta\phi}\int_{0}^{\Delta\phi}d\phi_1d\phi_2\left[\frac{V(\phi_1)+V(\phi_2)-V(\phi_1+\phi_2)}{\sqrt{V(\phi_1)V(\phi_2)}}\right]+\mathcal{O}[(\gamma v)^{-2}]~,\nonumber\\ 
\end{eqnarray}
where $M$ is the energy of the stationary kink.
\end{itemize}
We claim that the above expressions provide the close-form answer for the phase shift and velocity change (to leading order in $1/(\gamma v)$) for collisions between {\it any} pair of periodic kinks in (1+1) dimensions. Importantly, the phase shift can be calculated simply by an integral over the potential. We do not need a solution for the field profile or the detailed dynamics of the interaction during the collision.

In Sec. \ref{sec-ex} we show that these results are consistent with the exact results for the integrable Sine-Gordon model. We also show that they agree extremely well with our detailed numerical simulations in models that are far from being integrable. 
We discuss our results and possible future directions in Sec. \ref{sec-dis}. Finally, in Appendix \ref{sec-energy} we describe the possible usefulness of an optical theorem for our framework. We hope that this work will serve as a stepping stone to a more general scattering theory of solitary waves. 

%~~~~~~~~~~~~~~~~ General Perturbative expansion~~~~~~~~~~~~~~~~~~~~~~~~~~~~~~~~
%%%%%%%%%%%%%%%%%%%%%%%%%%%%%%%%%%%%%%%%%%%%%%%%
\section{General Perturbative Expansion}
\label{sec-GPertExp}
In this section we describe our general framework for understanding collisions of solitary waves.  We focus on solitary waves in scalar field theories, however the general idea is more widely applicable. We will discuss simplifying assumptions regarding the scalar field potential and the solitary waves under consideration. A large class of theories and their corresponding solitary wave solutions such as Q-balls\cite{Col85,Lee:1991ax}, oscillons \cite{Bogolyubsky:1976yu, Gleiser:1993pt, CopGle95, Amin:2013ika, Amin:2010jq}, kinks and domain walls are consistent with these assumptions. We will then show that the effect of collisions between ultrarelativistic solitary waves can be understood and calculated in a controlled fashion. 

We keep the discussion quite general in this section. The reader craving more concreteness can refer to Sec.\ref{sec-kinks} and Sec.\ref{sec-pert} where we apply our framework to (1+1) dimensional kinks. Some of the arguments below will be repeated there.

 %Discussion of non-linear perturbations is provided Appendix \ref{sec-nonlinear} as well as understanding of their long time behavior is disc.

%~~~~~~~~~~~~~~~~~figure Potential~~~~~~~~~~~~~~~~~~~~~~~~~~~~~~~~~~~~~~~~~~~~~~~
\begin{figure}
\begin{center}
\includegraphics[width=14cm]{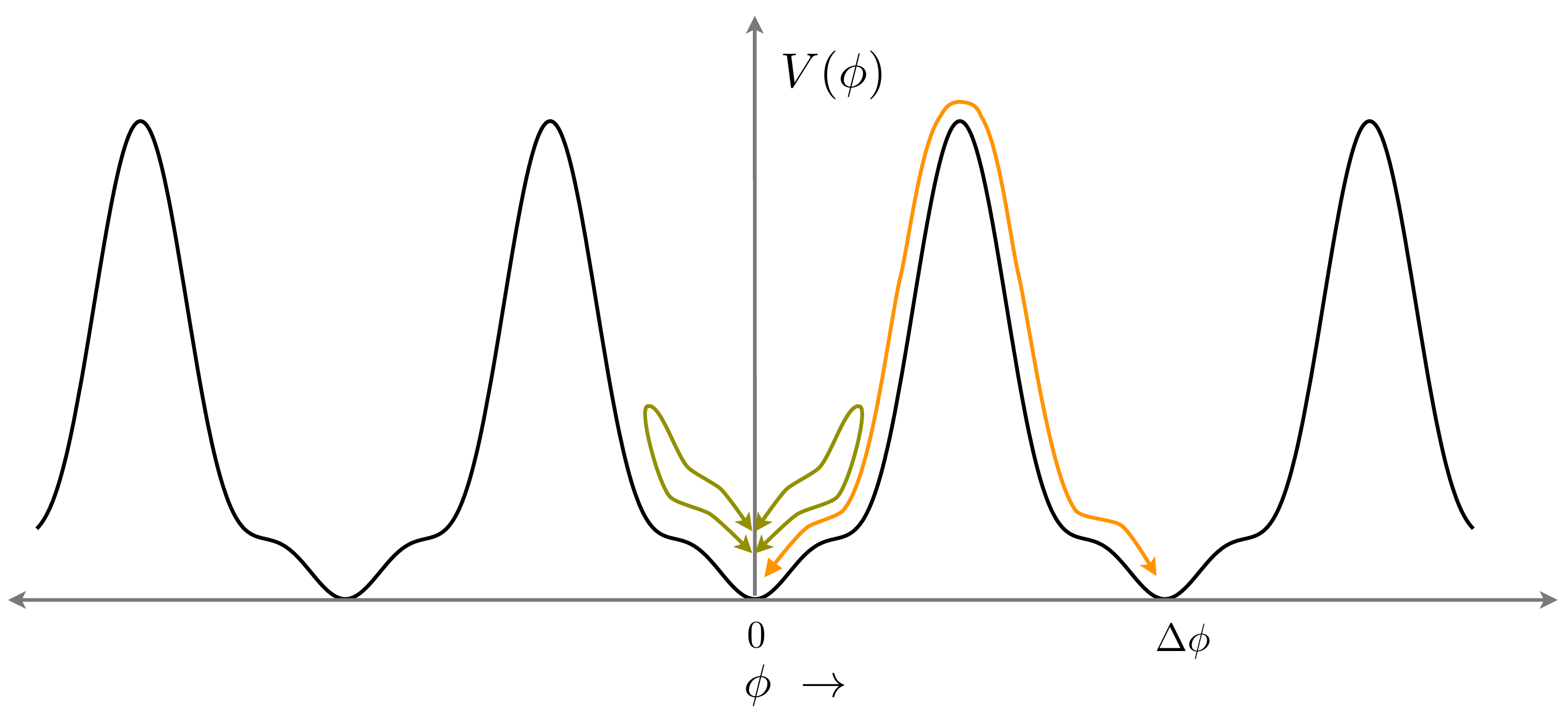}
\caption{Some localized solitary wave solutions probe the potential around a single minimum (for example, oscillons (green)). For others (such as kinks (orange)), the field configuration interpolates across the barrier between minima. For the latter case, our framework is applicable for potentials that are periodic in the field range probed by the two solitary wave solutions participating in the collision. 
\label{fig-Potential}}
\end{center}
\end{figure}
%~~~~~~~~~~~~~~~~~~~~~~~~~~~~~~~~~~~~~~~~~~~~~~~~~~~~~~~~~~~~~~~~~~~~~~

\subsection{Scalar Field Potential}
Consider a canonical scalar field with a Lagrangian
\begin{eqnarray}
\mathcal{L} &=& \frac{1}{2}(\partial_t\phi)^2-\frac{1}{2}(\nabla\phi)^2 -V(\phi)~.
\end{eqnarray}
The equation of motion is\footnote{We use the ``mostly plus" convention for the Minkowski metric.}
\begin{eqnarray}
& &\Box\phi=V'(\phi)~.
\label{eq-generalEOM}
\end{eqnarray}
We assume that the potential has a minimum (vacuum) at $\phi=0$ with $V(0)=0$. The potential can have multiple vacua as long as it is periodic: $V(\phi)=V(\phi+\Delta\phi)$. Therefore, every vacuum is effectively the same. An example of such a potential is shown in Fig.\ref{fig-Potential}. Our assumption regarding the potential is driven by practical considerations. We need a calculable background solution for perturbation theory to be easily applicable. For the potentials above, a linear superposition of two solitary waves provides a good background solution during collisions.\footnote{We verify this by calculating the correction to this superposition. The fact that superposition is a good solution before the collision comes from our assumption of localization discussed in the next subsection.} Thus if individual solitary wave solutions are available (analytically or numerically), a good background solution can be easily constructed. 

\begin{figure}
\begin{center}
\includegraphics[width=10cm]{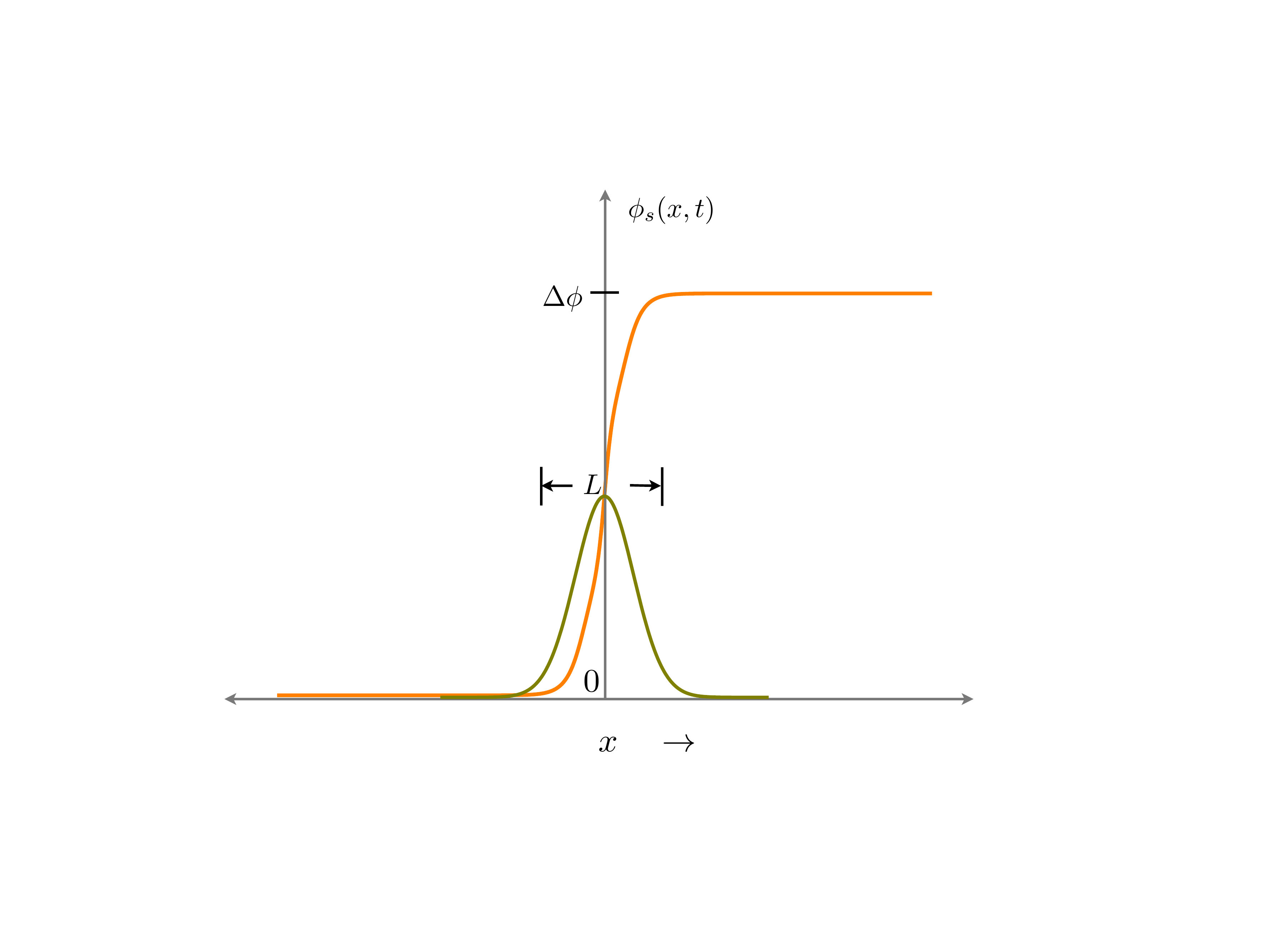}
\caption{Two examples of localized solitary waves. A kink solution (orange) interpolating between adjacent vacua as $x\rightarrow \pm \infty$. An oscillon solution (green) interpolating between the same vacuum at $x\rightarrow\pm \infty$. The oscillatory time dependence of the oscillon solution is not shown. In both cases, the field deviates away from the two vacuum over a length-scale $\sim L$.
\label{fig-profile}}
\end{center}
\end{figure}

While this restriction on the potential might appear severe, note that the potential only has to be periodic for the field range explored by the solitary wave solutions. For example, individual kinks probe adjacent minima whereas two kinks interacting during a collision can probe multiple nearby minima. In contrast, small amplitude oscillons only probe a single minimum of the potential. Examples of field profiles of these {\it individual} solitary waves are shown in Fig.\ref{fig-profile}.

For some non-periodic potentials (given the conditions specified in \cite{GibLam10}), a modification on top of the superposition is required to obtain a good background solution. If a good background solution is available, our perturbative framework may be generalized to include these cases. 

\subsection{Localized Solutions}

We require that the solitary waves under consideration be spatially localized along the direction of collision. Along the orthogonal direction we require the objects to be either localized or possess a symmetry that renders the infinite directions redundant. For example, domain walls often possess such a symmetry.

A solitary wave solution $\phi_s(\bx,t)$ is localized if for every $\epsilon$, there is a center $\bx_0$ and a size $L$ with $|\phi_s(\bx,t)-\phi_v|<\epsilon$ for all $|\bx-\bx_0|>L/2$.\footnote{Recall that it is fine to have spatial directions along which this criterion is not satisfied as long as there exists a symmetry that renders those directions redundant. The symmetry can be an approximate one. For example, as long as the radius of curvature of a domain wall is much larger than its thickness, it effectively has a planar symmetry.}  Here $\phi_v=N\Delta \phi$ with $N=0, \pm 1, ...$ is the vacuum value of the field. One can picture a localized solution as having a small tail that rapidly approaches a minimum of $V(\phi)$ beyond some length scale $\sim L$ away from its center.  
%------------------------------------------
%------------------------------------------

In this paper we assume exponentially suppressed tails. This has the advantage that along with the field profiles, other quantities derived from them (for example spatial integrals or derivatives of the profile) also have exponentially suppressed tails. In many cases, power law tails can be sufficient as well, but a more detailed statement has to be made about when they can be ignored. We leave that for future work.

Having discussed the localization criterion, let us now turn to the effects of a collision between such solitary waves.

%~~~~~~~~~~~~~~~~~figure Interaction~~~~~~~~~~~~~~~~~~~~~~~~~~~~~~~~~~~~~~~~~~~~~~~
\begin{figure}
\begin{center}
\includegraphics[width=15cm]{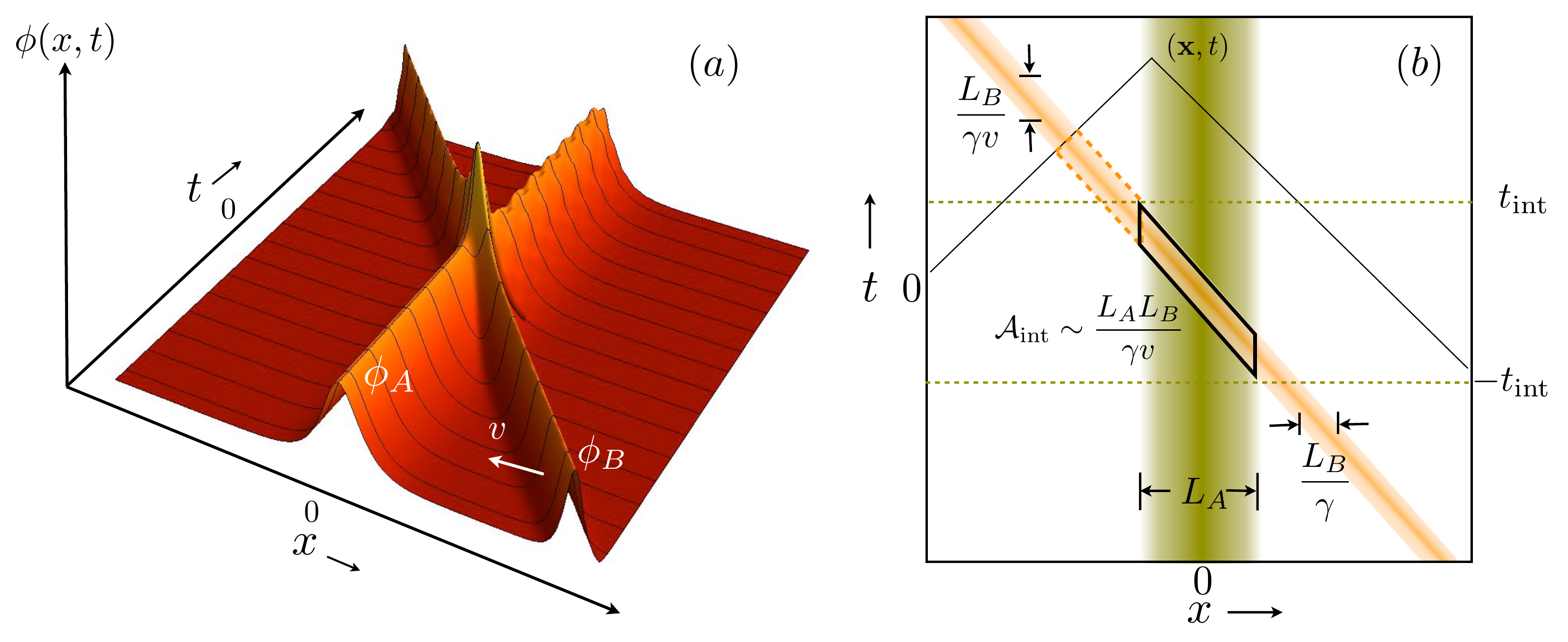}
\caption{(a) The figure shows a collision between a stationary solitary wave $\phi_A$ and a fast moving one $\phi_B$ as a function of space and time (not an actual simulation). Notice the Lorentz contraction in the spatial width of the fast moving soliton. Also note the small perturbations generated by the collision. (b) A simplified but useful view of the collision, highlighting the space-time area occupied by the solitary waves. The strips show the space-time area where the field deviates significantly from its vacuum value(s). The vertical (green) strip represents the space-time area occupied stationary solitary wave, whereas the thin diagonal strip (orange), represents the area occupied by a fast moving one. The effective overlap area, or the area of interaction $\Aint$ is shown as a black parallelogram in the middle. The time $t=\tint$ marks the end of the collision. When the incoming solitary wave is ultrarelativistic, the interaction area is suppressed by $1/(\gamma v)$.  \label{fig-Int}}
\end{center}
\end{figure}
%~~~~~~~~~~~~~~~~~~~~~~~~~~~~~~~~~~~~~~~~~~~~~~~~~~~~~~~~~~~~~~~~~~~~~~

\subsection{A Recursive Expansion}

Consider a stationary solitary wave $\phi_A(\bx,t)$ centered around $x=0$, and another solitary wave $\phi_B(\bx,t)$ that is moving toward $\phi_A$ from far away with its center at $x=-vt$ and $\gamma=1/\sqrt{1-v^2}\gg 1$. We assume that they are localized along the direction of collision with extents $L_A$ and $L_B$ respectively in their own rest frames. A collision of such solitary waves is shown heuristically in Figure \ref{fig-Int}(a). Note that the spatial extent of the fast moving solitary wave: $L_B/\gamma$ is Lorentz contracted in the rest of frame of $\phi_A$. This can be seen more clearly in Figure \ref{fig-Int}(b) where we simply show the space-time area where their field values deviate significantly from vacuum (or different vacua). Note that in both figures, the directions orthogonal to the collision are not shown.  

If $(\phi_A+\phi_B)$ is a good approximation before, during, and after the collision, then all physical effects of a collision happen only when the two solutions overlap in the {\it interaction} area (see the outlined parallelogram in Figure \ref{fig-Int}(b)):
\begin{eqnarray}
\Aint\sim \frac{L_AL_B}{(\gamma v)}~,
\label{eq:Aint}
\end{eqnarray}
where again the orthogonal dimensions are suppressed. This $1/(\gamma v)$ suppression of the interaction area plays the role of the perturbation parameter in analyzing highly relativistic collisions. 

\begin{figure}
\begin{center}
\includegraphics[width=15cm]{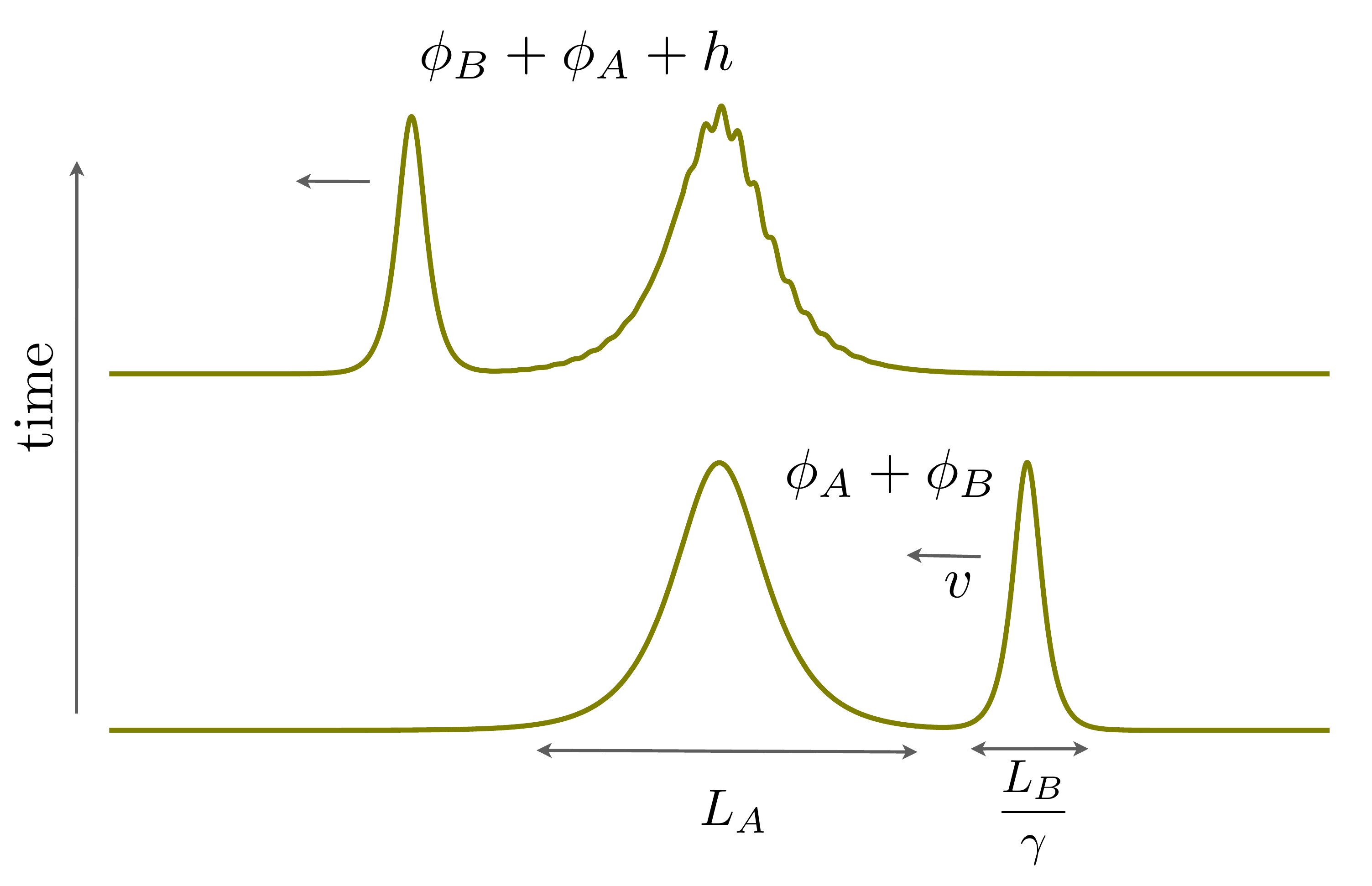}
\caption{The field profile before the collision (bottom), and the profile after the collision (top).
\label{fig-beforeafter}}
\end{center}
\end{figure}

Before the two solitary waves overlap, their linear superposition $(\phi_A+\phi_B)$ satisfies equation (\ref{eq-generalEOM}) if we ignore the exponentially suppressed tails (see lower plot in Fig. \ref{fig-beforeafter}). {\it Our claim is that $(\phi_A+\phi_B)$ continues to provide a good approximation even after the collision. We justify our claim by calculating the corrections to it and showing that they are indeed small. We calculate the corrections using a perturbative framework that uses $1/(\gamma v)$ as the small parameter.} 

First, we write the general solution as
\begin{eqnarray}
\phi(\bx,t)=\phi_A(\bx,t)+\phi_B(\bx,t)+h(\bx,t)~.
\end{eqnarray}
The linearized equation of motion for $h$ is\footnote{The nonlinear effects from higher order terms in $h$ are discussed in the Appendix \ref{sec-nonlinear}.}
\begin{equation}
\left[\Box -W_0(\bx,t)\right]h = S(\bx,t)+\Delta W(\bx,t)h~,
\label{eq-generalEOMh}
\end{equation}
where for future convenience we have defined 
\begin{eqnarray}
S(\bx,t) &\equiv& V'(\phi_A+\phi_B)-V'(\phi_A)-V'(\phi_B)~,\\
W_0(\bx,t) &\equiv& V''(\phi_A)~,\\
\Delta W(\bx,t) &\equiv& V''(\phi_A+\phi_B)-V''(\phi_A)~.
\label{eq-generalSource}
\end{eqnarray}
We refer to $S$ as the source, $W_0$ as the mass term (related to the stationary solitary wave) and $\Delta W$ as the change in mass due to the fast moving solitary wave. Note that $W_0$ is nonzero in the space-time area occupied by $\phi_A$ (the entire green strip in Fig. \ref{fig-Int}(b)) and $\Delta W$ is nonzero in the space-time area occupied by $\phi_B$ (see orange strip in Fig. \ref{fig-Int}(b)). However, $S$ is nonzero only in the overlap region $\Aint$. Moreover, the maximum/minimum values of $S$, $W_0$ and $\Delta W$ do not depend on $v$, a direct consequence of dealing with a Lorentz invariant scalar field theory.
%~~~~~~~~~~~~~~~~~figure BeforeAfter~~~~~~~~~~~~~~~~~~~~~~~~~~~~~~~~~~~~~~~~~~~~~~~
%~~~~~~~~~~~~~~~~~~~~~~~~~~~~~~~~~~~~~~~~~~~~~~~~~~~~~~~~~~~~~~~~~~~~~~
Our definition of $W_0$ in terms of $\phi_A$ and the way we have arranged equation (\ref{eq-generalEOMh}) reflects our choice to concentrate on the the effects of the collision on $\phi_A$.\footnote{This focus on the stationary solitary wave is partially driven by the fact that in the special case where the solitary waves are static in their rest frame, their perturbations can be expanded in terms of a separable eigenmode basis.} We concentrate on evaluating $h(\bx,t)$ in the region $|x|\lesssim L_A/2$ and $t \sim \tint$ where $\tint\equiv v^{-1}(L_A+\gamma^{-1}L_B)$ marks the end of the collision (upper dashed line in Figure \ref{fig-Int}(b)).  

Let us first see why $h(\bx,t)$ is expected to be small in ultrarelativistic collisions. Before the collision begins, i.e. for $t<-\tint$, $h(\bx,t)=0$ is a consistent solution of equation (\ref{eq-generalEOMh}). Then the solution for $t\sim \tint$ is 
\begin{eqnarray}
h(\bx,t)\sim \int dt' d\bx' \mG(\bx,t;\bx',t')S(\bx',t')\sim \frac{1}{(\gamma v)}~,
\end{eqnarray}
where $\mG$ is a Green's function satisfying 
\Beq
\left[\Box-W_0(\bx,t)\right] \mG(\bx,t;\bx',t')=\delta(\bx-\bx',t-t')~. \label{eq-Green}
\Eeq
To understand the above expression for $h$ and the scaling with $(\gamma v)$ recall that $S$ is nonzero only when the solitary waves overlap, i.e. in the interaction area $\Aint\propto 1/(\gamma v)$ and that the magnitude of $S$ does not depend on $\gamma v$ in this interaction area. These two properties of $S$ yields the desired scaling of $h\sim 1/(\gamma v)$. Note that in the above discussion we have ignored the $\Delta W h$ term because this term becomes non-zero only after $h$ becomes non-zero, and will hence be a higher order effect. 

If we wish to go beyond this leading order description, it is convenient to expand $h$ as a recursive series:
\begin{eqnarray}
h(\bx,t)=\sum _{n=1}h^{(n)}(\bx,t)~.
\end{eqnarray}
The $h^{(n)}$ can be obtained by recursively solving \eqn{eq-generalEOMh} using the the Green's function $\mG$ defined in \eqn{eq-Green}:
\begin{eqnarray}
h^{(1)}(\bx,t)&=&\int dt' d\bx' \mG(\bx,t;\bx',t')S(\bx',t')~,\\
h^{(n)}(\bx,t)
&=&\int dt' d\bx'\mG(\bx,t;\bx',t') \Delta W(\bx',t')h^{(n-1)}\qquad n\ge 2~.
\end{eqnarray}
As stated earlier, $S$ is only active in the interaction area $\Aint$. $\Delta W h$ will be active in a larger (presumably infinite) area, but only contributes to the integral below $t$ because of the Green's function. For $t\sim \tint$, this contribution comes from the dashed orange parallelogram along with the solid black one in Fig. \ref{fig-Int}(b). This active area is similarly suppressed, $\Aint'\sim \Aint \propto 1/(\gamma v)$. Thus, if we only care about the value of $h$ up to time $t\sim\tint$, our recursive expansion $h^{(n)}$ has the following convenient scaling.
\begin{eqnarray}
h^{(1)}(\bx,t)&=&\int_{\Aint} dt' d\bx' \mG(\bx,t;\bx',t')S(\bx',t')\sim \frac{1}{(\gamma v)}~, \label{eq-hSource}\\
h^{(n)}(\bx,t)
&=&\int_{\Aint'} dt' d\bx'\mG(\bx,t;\bx',t') \Delta W(\bx',t')h^{(n-1)}\sim \frac{1}{(\gamma v)^n}\qquad n\ge 2
\label{eq-generalRecur} 
\end{eqnarray}
From the above expressions, it is easy to see that at $t\sim \tint$ and $(\gamma v)\gg 1$: $h^{(n)}\propto \Aint h^{(n-1)}\sim 1/(\gamma v)^n$. The long term behavior of the perturbations, $h(t\gg\tint)$ is a bit more subtle and is discussed in Appendix \ref{sec-LongTime}. 

In summary, from the point of view of perturbations on top of the solitary wave $\phi_A$, the ``collision'' is nothing but an ``overlap'' of the background. The overlap results in some temporary nontrivial dynamics. When $(\gamma v)\gg1$, the nontrivial dynamics lasts a short time and happens in a small spatial region. Thus, it leaves a small effect when the incoming solitary wave leaves.\footnote{This is reminiscent of the sudden collision approximation in heavy ion collisions. We thank David Seery and Cliff Burgess for pointing this out to us. A large $\gamma$ limit has also been used in gravitational self-force calculations \cite{Galley:2013eba}.} We draw that pictorially in Fig \ref{fig-beforeafter}. In the upcoming sections we will calculate $h$ for a collision between a stationary and incoming kink. Although not evident from the notation, we will always assume that we are evaluating the perturbation in the vicinity of the stationary solitary wave and in the time interval $t\sim \tint$.

%~~~~~~~~~~~~~~~~ 1+1 Dimensional Kinks ~~~~~~~~~~~~~~~~~~~~~~~~~~~~~~~~~~~~~~~
%%%%%%%%%%%%%%%%%%%%%%%%%%%%%%%%%%%%%%%%%%%%%%%%

\section{(1+1) Dimensional Kinks}
\label{sec-kinks}

In this section we discuss an example of solitary waves, namely isolated kinks in (1+1) dimensional scalar field theories. We will study their collisions in Sec. \ref{sec-pert}.  First, we review general properties of individual kinks. We then show why they can be considered localized in the sense discussed in Sec. \ref{sec-GPertExp}, which in turn allows us to apply our general framework to study their collision. We chose (1+1) dimensional kinks because their properties are well understood. Moreover, unlike intrinsically time dependent solitary waves like oscillons, stationary kinks are time independent. This allows us to decompose linear perturbations around the kinks into a convenient eigenmode basis, which we use in calculating and expressing the physical meaning of $h^{(i)}$.

To avoid possible confusion in the upcoming sections, we pause briefly to clarify our notation. Primes $'$ on a function with a single argument denote partial derivative with respect to that argument, so $f'(u) = \partial_u f$ where $u=\phi,x,t$ or $\gamma(x+vt)$ (for example). We will always denote the argument when using the prime notation. When there is an ambiguity, we will restore the partial derivatives.

\subsection{Background}

Consider a 1+1 dimensional, canonical scalar field theory with a periodic potential $V(\phi)=V(\phi+\Delta \phi)$ (see Figure \ref{fig-Potential}). The equation of motion for the scalar field is given by 
\begin{eqnarray}
\partial_t^2\phi(x,t)-\partial_x^2\phi(x,t)+V'[\phi(x,t)]=0~.
\label{eq-EOM}
\end{eqnarray}
A classic (though somewhat special) example of such a scalar field theory is the Sine-Gordon theory with $V(\phi)=m^2(1-\cos \phi)$ and $\Delta\phi=2\pi$. Note that in (1+1) dimensions, $\phi$ is dimensionless, and $V(\phi)$ has the dimensions of (energy)$^2$.  We assume our potential has degenerate minima at $\phi=N\Delta\phi$ for integer $N$ with $V(N\Delta\phi)=0$. We also assume that the minima are quadratic and convex, that is  $V''(N\Delta \phi)=m^2>0$. 

In scalar field theories with periodic potentials, there exist static, minimum-energy field configurations called kinks: $\phi_K(x)$. In such configurations, the field values interpolate between neighboring minima of the potential. Taking the field values at the  two minima to be $0$ and $\Delta\phi$, such a kink satisfies the relations
 \begin{eqnarray}
\phi_K''(x) &=& V'[\phi_K(x)]~, \label{eq-xeom} \\
\lim_{x\rightarrow-\infty}\phi_K(x) &=& 0~, \\
\lim_{x\rightarrow\infty}\phi_K(x) &=& \Delta\phi~. 
\end{eqnarray}

\begin{figure}
\begin{center}
\includegraphics[width=10cm]{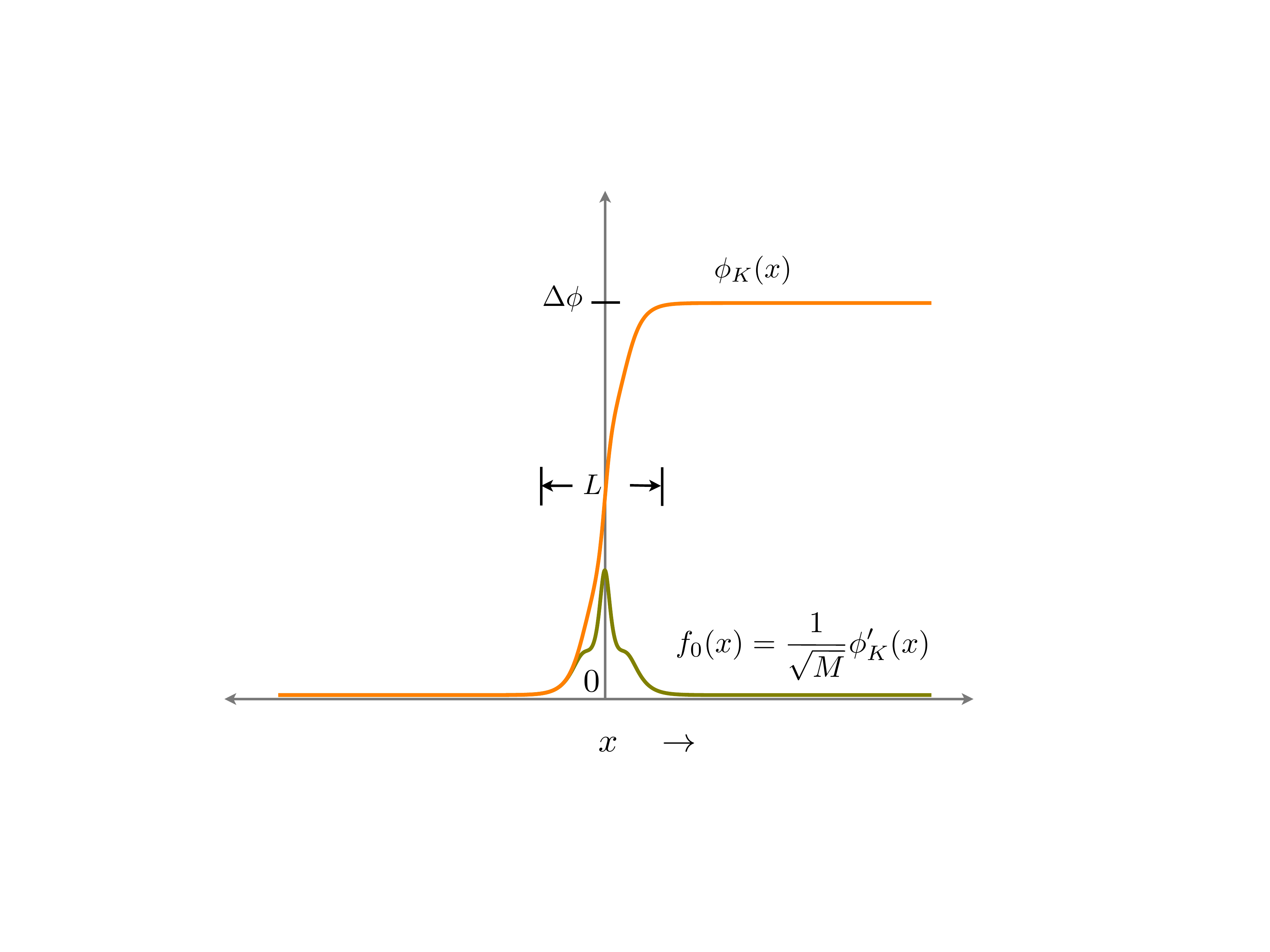}
\caption{The orange curve is the field profile of the kink for the potential given in Fig.1. We can treat it as an object of finite size $L$, since away from this central region the field profile exponentially approaches the vacuum value, $0$ or $\Delta\phi$. The green curve is the normalized spatial derivative of the profile.
\label{fig-Kink}}
\end{center}
\end{figure}

In general, the field profile changes significantly only in an interval of length $\sim L$ around the center of the kink (see Figure \ref{fig-Kink}). Thereafter, the field decays exponentially because of the mass term $V''(N\Delta \phi)=m^2>0$: 
\begin{eqnarray}
\phi_K(x) \,\,{\rm mod}\,\, \Delta\phi&\sim e^{-m|x|}\lesssim e^{-m L/2}~,\quad\quad\quad {\rm{for}}\quad|x|>L/2 \label{eq-center}~.
\end{eqnarray}

%~~~~~~~~~~~~~~~~~~~~~~~~~~~~~~~~~~~~~~~~~~~~~~~~~~~~~~~~~~~~~~~~~~~~~~
%~~~~~~~~~~~~~~~~~~~~~~~~~~~~~~~~~~~~~~~~~~~~~~~~~~~~~~~~~~~~~~~~~~~~~~

Such field configurations, where the field deviates from its vacuum values in a finite region $L$ and thereafter decays exponential is precisely the kind of objects where our general framework of Sec. \ref{sec-GPertExp} is applicable. As discussed in that section, after picking an $L$ we will ignore the tails beyond it. Explicitly, we make the following pragmatic assumption which significantly simplifies our analysis:
\begin{equation}
\phi_K(x) \,\,{\rm mod}\,\, \Delta\phi=0\quad\quad\quad{\rm{for}}\quad |x|>L/2~. 
\label{eq-prac}
\end{equation}  
As we are primarily interested in investigating the effects of kink-kink and kink-antikink collisions where short-range interactions dominate, the long range interaction arising from exponentially suppressed tails can be safely ignored. For the interested reader, the effects arising from the exponentially suppressed tails have been studied in the literature, notably in \cite{Manton:1978gf}. 

The total energy (sometimes colloquially called mass -- not to be mistaken with the mass $V''(\phi)$ term of a potential) of a stationary kink is  
\begin{eqnarray}
M = \int_{-\infty}^{\infty} dx
\left(\frac{1}{2}\phi_K'(x)^2 + V[\phi_K(x)] \right)= \int_0^{\Delta\phi} d\phi ~ \sqrt{2V(\phi)}~.\label{eq-mass}
\end{eqnarray}
We have used $\phi_K'(x)=\sqrt{2 V[\phi_K(x)]}$, which can be obtained by integrating the equation of motion $\phi_K''(x)=V'[\phi_K(x)]$ by parts. Ignoring the exponentially damped tails, $M=\int_{-L/2}^{L/2} dx~\left(\frac{1}{2}\phi_K'(x)^2 + V[\phi_K(x)] \right).$

Now, let us consider a kink moving at a constant speed $v$ from the positive to negative direction. Since our theory is Lorentz invariant, such a moving kink is simply given by boosting the stationary kink profile:
\begin{equation}
\phi(x,t) = \phi_K\left[\gamma(x+vt)\right]~,
\label{eq-boosted}
\end{equation}
where $\gamma= (1-v^2)^{-1/2}$ is the Lorentz boost factor. The total energy of this moving kink is
\begin{eqnarray}
\int_{-\infty}^{\infty}dx\left(\frac{1}{2}(\partial_t \phi_K[\gamma(x+vt)])^2+\frac{1}{2}(\partial_x\phi_K[\gamma(x+vt)])^2+V(\phi_K[\gamma(x+vt)])\right)=\gamma M~. \nonumber \\
\end{eqnarray}
Hence its energy behaves like the energy of a point particle. 

\subsection{Perturbations around an Isolated Kink}

Before moving on to collisions, we discuss linearized perturbations around isolated kinks and decompose them into a convenient eigenmode basis. As noted before, this eigenmode basis will simplify calculations and allow for a useful physical interpretation of the perturbations. Consider a small perturbation on top of a single static kink,
\begin{eqnarray}
\phi(x,t)=\phi_K(x) + h(x,t)~.
\label{eq-solCoM}
\end{eqnarray}
It obeys the equation of motion (at linear order in $h$)
\begin{equation}
\partial_t^2 h - \partial_x^2 h + W_0(x) h = 0~,
\label{eq-hSingle}
\end{equation}
where
\begin{equation}
W_0(x) \equiv V''[\phi_K(x)]~.
\label{eq-W0}
\end{equation}
The general solution can be decomposed into discrete (localized) $f_i(x)$ and continuous (free) $f_w(x)$ eigenmodes
\begin{eqnarray}
h(x,t) &=& \sum_i g_i(t) f_i(x) + \int_{w_c}^\infty g_w(t) f_w(x) dw ~,
\nonumber \\
&\equiv& \sum_{a=0}^{\infty} g_a(t) f_a(x)~.
\label{eq-modes}
\end{eqnarray}
In the second line above, we have abbreviated the integral over the continuous modes\footnote{Note that the dimensionality of the modes $f_w(x)$ and $f_i(x)$ are different.} and written everything as a discrete sum.  As usual, both continuous and discrete eigenmodes $f_a(x)$ are the solutions of the eigenvalue problem
\begin{equation}
\left[-\partial_x^2 +W_0(x)\right]f_a(x)=E_af_a(x)~\label{eq-timeInd},
\end{equation}
and form a complete orthonormal basis labeled by $``a"$ in equation (\ref{eq-modes}). 
\begin{eqnarray}
\int_{-\infty}^\infty f_i(x) f_j(x) dx &=& \delta_{ij}~, \label{eq-ortho} \\
\int_{-\infty}^\infty f_{w_1}(x) f_{w_2}(x) dx &=& \delta(w_1-w_2)~, \\
\int_{-\infty}^\infty f_i(x) f_{w}(x) dx &=& 0~.
\end{eqnarray}
Plugging equation (\ref{eq-modes}) into equation (\ref{eq-hSingle}), and using the orthonormality of $\{f_a(x)\}$ we have an equation of motion for $g_a$:
\begin{eqnarray}
g_a''(t)+E_ag_a(t)=0~. \label{eq-ga}
\end{eqnarray}
Since $W_0(x)$ is time-independent, these eigenmodes represent ``stationary'' excitations on top of a single kink akin to the stationary states of a time-independent Sch\"{o}dinger's Equation. Hence it is convenient to express any small perturbation as a sum of eigenmodes on top of the kink. 

The eigenmode with the smallest eigenvalue is the {\em zero mode} with eigenvalue $E_0=0$. As we will show below, the $E_0=0$ mode is associated with the boost and translational symmetry of the original equation of motion -- the action of the zero mode is to shift the {\em phase} and the {\em velocity} of the kink. The normalized eigenmode that solves equation (\ref{eq-timeInd}) with $E_0=0$ is given by
\begin{eqnarray}
f_0(x)&=&\frac{1}{\sqrt{M}}\phi_K'(x)~\label{eq-f0def}, 
\end{eqnarray}
where $M$ is the mass of the kink. This can be checked immediately using equations (\ref{eq-xeom}) and (\ref{eq-mass}). The time dependence of the zero mode can be obtained from equation (\ref{eq-ga}) as 
\begin{eqnarray}
g_0(t)=A_0+B_0t~\label{eq-g0def}.
\end{eqnarray}
To understand the physical meaning of $A_0$ and $B_0$, consider a small, spatial translation $(\Delta x)$ of the stationary kink profile: 
\begin{eqnarray}
\phi_K[x-(\Delta x)]
&=&\phi_K(x)-\phi_K'(x)(\Delta x)+{\cal O}(\Delta x)^2+\dots,\nonumber\\
&\approx&\phi_K(x)-\sqrt{M}f_0(x)(\Delta x)~\label{eq-xshift}.
\end{eqnarray}
Similarly, a small velocity perturbation $(\Delta v)$ of the stationary kink profile yields (with $\gamma = (1-\Delta v^2)^{-1/2}\approx 1+(\Delta v)^2/2+\dots$)
\begin{eqnarray}
\phi_K[\gamma(x-(\Delta v)t)]
&=&\phi_K(x)-\phi_K'(x)(\Delta v) t+{\cal O}(\Delta v)^2+\dots~,\nonumber\\
&\approx&\phi_K(x)-\sqrt{M}f_0(x)(\Delta v) t~.\label{eq-vshift}
\end{eqnarray} 
For comparison, consider the eigenmode expansion of the solution:
\begin{eqnarray}
\phi(x,t)
&=&\phi_K(x)+h(x,t)~,\nonumber\\
&=&\phi_K(x)+g_0(t)f_0(x)+\sum_{a=1}g_a(t)f_a(x)+...~,\nonumber\\
&=& \phi_K(x)+(A_0+B_0 t)f_0(x)+\sum_{a=1}g_a(t)f_a(x)+...~.
\end{eqnarray}
Comparing equation (\ref{eq-xshift}) and (\ref{eq-vshift}) to the mode expansion above, we see that
\begin{eqnarray}
(\Delta x)&=&-\frac{1}{\sqrt{M}}A_0~,\label{eq-xA0shift}\\
(\Delta v)&=&-\frac{1}{\sqrt{M}}B_0~.\label{eq-vB0shift}
\end{eqnarray}
As promised, the coefficients $A_0$ and $B_0$, which characterize the amplitude and evolution of the zero mode (see equation (\ref{eq-g0def})), determine the phase shift and velocity change of the solitary wave. Arguably, the phase shift and velocity change are the most important physical outcomes of ultrarelativistic collisions. We will calculate their leading order values (part of a series in $(\gamma v)^{-1}$) in the upcoming sections. The other modes with $E>0$ are related to oscillating fluctuations on top of the kink. 

Note that the zero mode expresses the Lorentz symmetry of the background solution. At leading order, we are free to describe a kink as a translated kink plus a compensating zero mode. However, for the convenience of the perturbative calculation we are about to do, we will choose the initial amplitude of the zero mode to be zero. The zero mode introduces further subtleties at higher order (for us third order in $1/(\gamma v)$), which can possibly be addressed by using collective co-ordinates \cite{Wei12}.\footnote{We thank Erick Weinberg for alerting us to this possibility.} We leave this for future work.

%~~~~~~~~~~~~~~~~~~ Perturbation Theory in $1+1$ dimensions ~~~~~~~~~~~~~~~~~~~~~~~~~~~~~~~~~~~~~~~~~~~~~~~~
%%%%%%%%%%%%%%%%%%%%%%%%%%%%%%%%%%%%%%%%%%%%%%%%%%%%%%%%%%%%%%%%%%

\section{Perturbation Theory for Kink Collisions in $1+1$ Dimensions}
\label{sec-pert}
\begin{figure}
\begin{center}
\includegraphics[width=14cm]{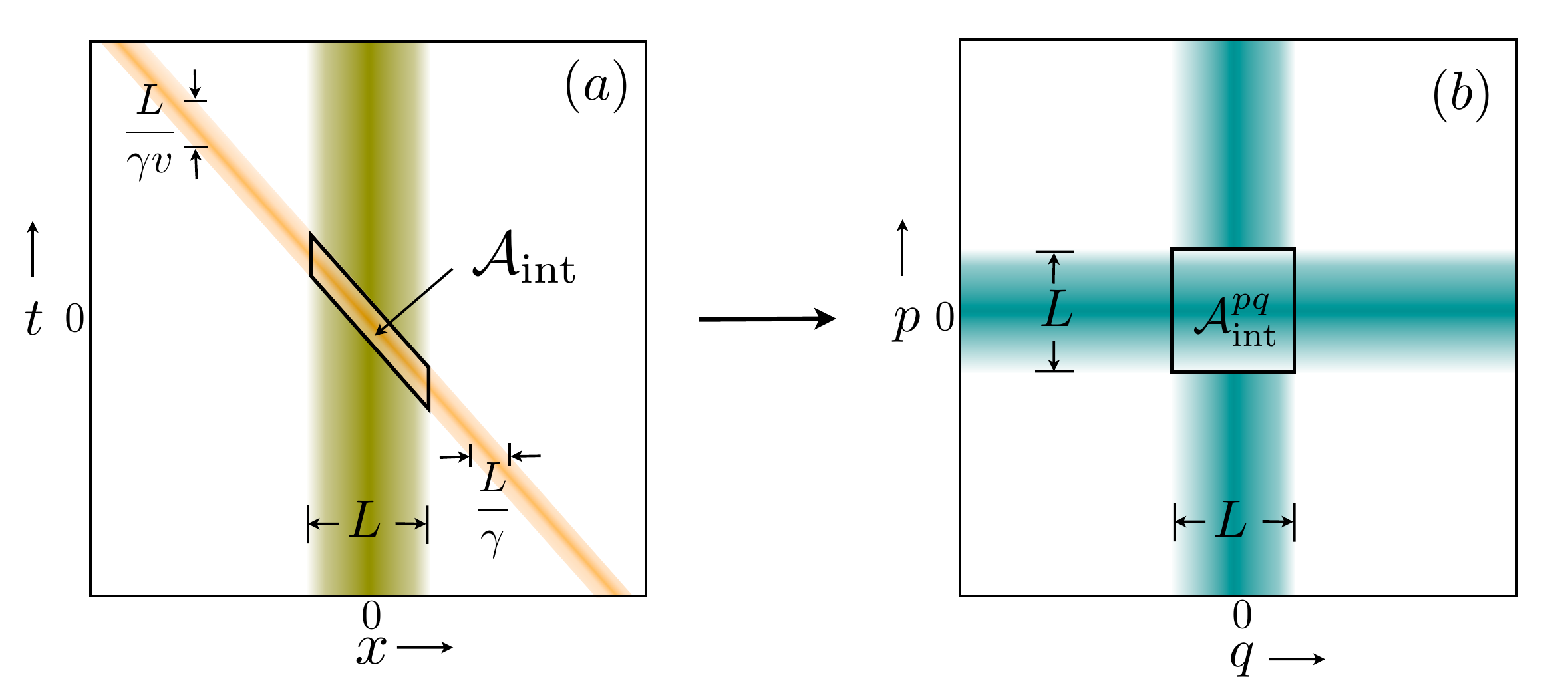}
\caption{The two strips represent the two colliding objects. The source function is only nonzero in the spacetime region $\Aint$ where the strips overlap with each other. Figure (a) is in the rest frame of one kink and it shows that the area is suppressed by $\gamma^{-1}$. Figure (b)  is in a convenient coordinate system we use to integrate over the effect of the source term.
\label{fig-Int123}}
\end{center}
\end{figure}

\begin{figure}
\begin{center}
\includegraphics[width=10.6cm]{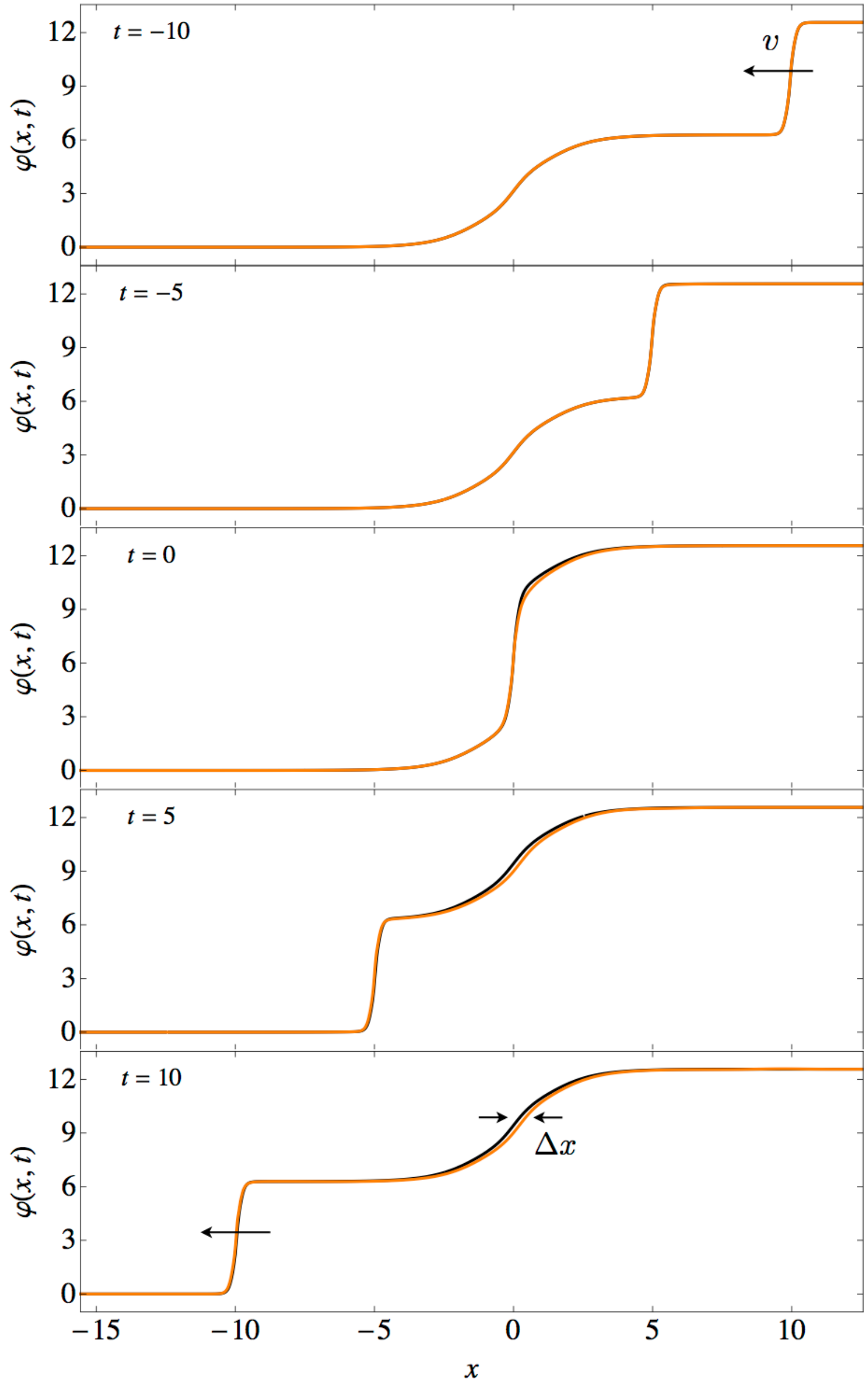}
\caption{The orange curves are the numerical field profiles before, during, and after collision. The black curve is the superposition solution which ignores all interactions. We can see a clear phase shift after the collision.
\label{fig-FieldPlot}}
\end{center}
\end{figure}

Consider a stationary kink $\phi_K(x)$ centered around $x=0$, and incoming kink $\phi_K[\gamma(x+vt)]$ approaching the stationary kink from the positive $x$ direction with a speed $v$ (see the top panel in Fig. \ref{fig-FieldPlot}). The kinks collide around $t=0$. The space-time area occupied by these kinks in the rest frame of the stationary kink as well as their interaction area is shown in Fig. \ref{fig-Int123}(a). The orange strip represents the fast moving solitary wave, whereas the green strip represents the stationary one. The interaction area $\Aint$ is denoted by the black parallelogram. Due to Lorentz contraction, the spatial width of the orange strip scales as $\gamma^{-1}$. The collision takes place during a time interval $|t|=\tint=(L/v)(1+\gamma^{-1})$ and the space time interaction area $\Aint\approx L^2/(\gamma v)$.

Now let us write down the full solution, before, during and after the collision as follows:\footnote{If the potential is not periodic, we have to subtract a ``reference'' $\Delta \phi$, i.e.
\begin{equation}
\phi(x,t)=\phi_K(x) + \phi_K[\gamma(x+vt)] - \Delta \phi +h(x,t).
\end{equation}
For more information on this requirement, see \cite{GibLam10}.
}
\begin{eqnarray}
\phi(x,t)=\phi_K(x) + \phi_K[\gamma(x+vt)] + h(x,t)~. \label{eq-2kink}
\end{eqnarray}
Using the equation of motion (\ref{eq-EOM}) and (\ref{eq-xeom}),  the linearized equation of motion for the perturbation $h(x,t)$ is
\begin{equation}
\partial_t^2h-\partial_x^2h +W_0(x)h=-\Delta W(x,t)h-S(x,t)~. \label{eq-master}
\end{equation}
where
\begin{eqnarray}
W_0(x)&\equiv&V''[\phi_K(x)]~, \\
\Delta W(x,t)&\equiv& V''(\phi_K(x)+\phi_K[\gamma(x+vt)])-W_0(x)~,\label{eq-dW}\\
S(x,t) &\equiv& V'(\phi_K(x)+\phi_K[\gamma(x+vt)])
-V'[\phi_K(x)]-V'[\phi_K[\gamma(x+v t)]~.\label{eq-source}
\end{eqnarray}
$W_0(x)$ is the mass term for the single, stationary kink. $\Delta W$ is the change in mass due to the fast moving kink and is non-zero only when evaluated on the moving kink, wheres $S$ is the source which is non-zero only when the two kinks overlap. Typical functional forms of these three quantities are shown in Fig. \ref{fig-SW}. Notice the Lorentz contraction of the spatial extent of $S$ and $\Delta W$. 

Now,  if we were to only consider perturbations about an isolated kink and ignored the second kink, the right hand side of equation (\ref{eq-master}) would be zero. However for the case under consideration, the right hand side of equation (\ref{eq-master}) gets two additional terms because of the second kink: an $h$ dependent term with coupling $\Delta W(x,t)$ and a $h$ independent external source term $S(x,t)$. Both $S$ and $\Delta W h$ only become active once the collision begins. While $\Delta W\ne 0$ before the collision, recall that a linear sum of two kinks is a solution before the collision and hence $h=0$ then.  In this sense we can think of the incoming kink, via its interaction with the stationary one, as sourcing perturbations of the stationary kink.  

We now want to see how the collision excites eigenmodes of the single kink $\{f_a(x)\}$  defined in \eqn{eq-timeInd}. First we write the general solution to equation (\ref{eq-master}) in terms of these $\{f_a(x)\}$: 
%~~~~~~~~~~~~~~~~~~~~~~~~~~~figure S and Delta W ~~~~~~~~~~~~~~~~~~~~~~~~~~
\begin{figure}
\begin{center}
\includegraphics[width=14cm]{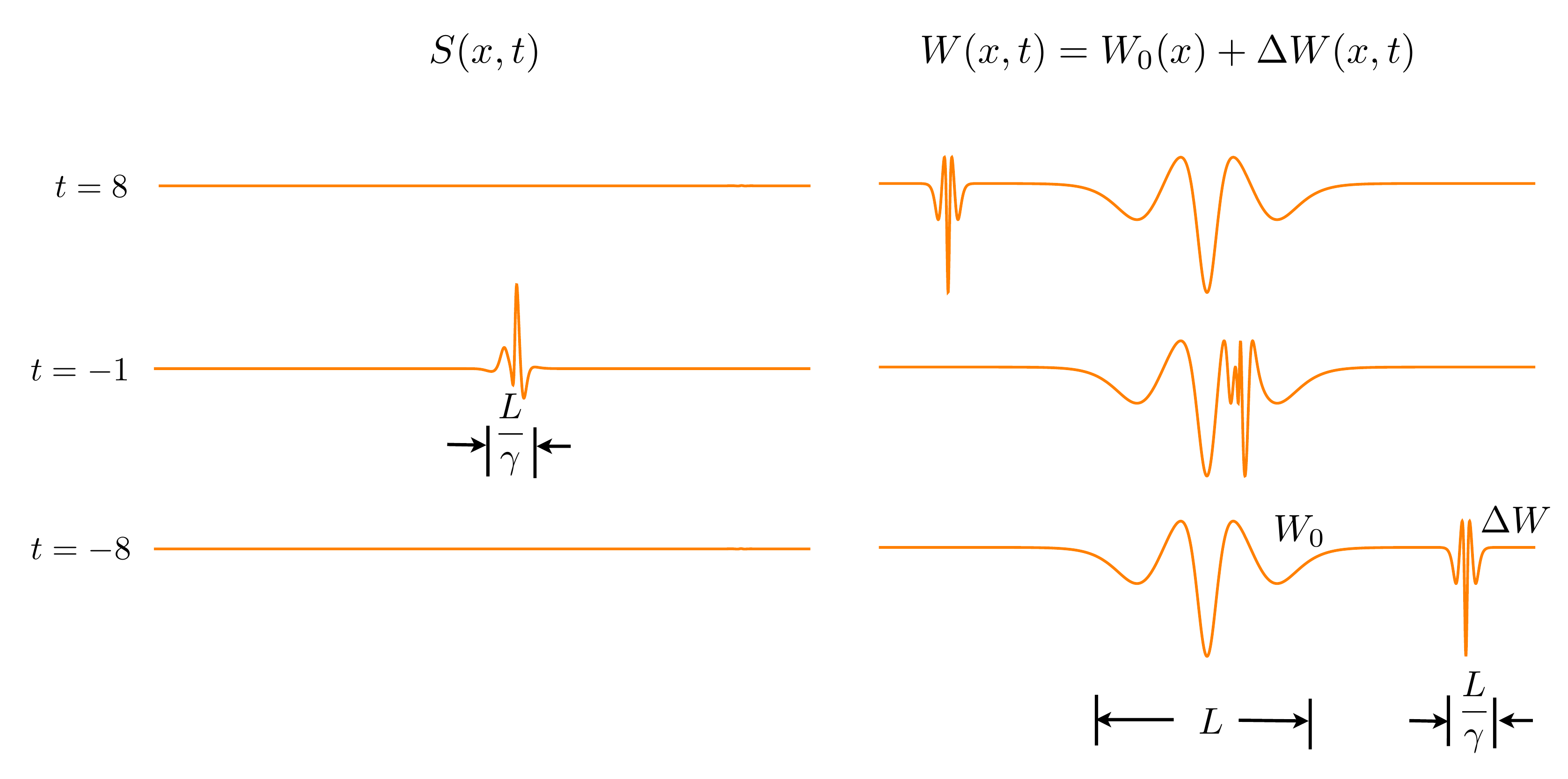}
\caption{In (a) we show snapshots of the source function $S(x,t)$ before, during and after the collision. $S(x,t)$ is non zero only during the collision and it's spatial extent at a fixed time is $L/\gamma$. In (b) we show $W(x,t)=W_0(x)+\Delta W(x,t)$. $\Delta W$ remains non-zero around the incoming kink and its spatial extent is also $L/\gamma$. We used the periodic $V(\phi)=(1-\cos \phi)(1-\alpha \sin^2\phi)$ with $\alpha = 0.7$ for these plots.
\label{fig-SW}}
\end{center}
\end{figure}
%~~~~~~~~~~~~~~~~~~~~~~~~~~~~~~~~~~~~~~~~~~~~~~~~~~~~~~~~~~~~~~~~~~~~~~
\begin{eqnarray}
h(x,t) = \sum_{a=0}^{\infty} G_a(t) f_a(x)~.
\label{eq-modesG}
\end{eqnarray}
We have used $G_a(t)$ instead of the free field $g_a(t)$ introduced in \eqn{eq-ga} (Sec. \ref{sec-kinks}) as the excitations are now sourced by the interaction terms $S$ and $\Delta W$. Plugging equation (\ref{eq-modesG}) into equation (\ref{eq-master}) and using the orthonormality of $f_a(x)$, we get

\begin{equation}
G_a''(t) +E_a G_a(t) =- \sum_bM_{ab}(t)G_b(t)-S_a(t)~\label{eq-Ga},
\end{equation}
where we have defined the transition matrix $M_{ab}(t)$ and the projected source $S_a(t)$ as:
\begin{eqnarray}
M_{ab}(t)&\equiv&\int dx~f_a(x) \Delta W(x,t) f_b(x)~, 
\label{eq-Mab}\\
S_a(t)&\equiv&\int dx~f_a(x)S(x,t)~\label{eq-Sa}. 
\end{eqnarray}
Recall that in the previous section,  \eqn{eq-ga} characterized the evolution of eigenmodes around an isolated kink. There was neither mixing of modes nor external sources. In  \eqn{eq-Ga} above, we see that a $G_a$ is sourced by $S_a$, and its evolution is a-priori coupled with all $G_b$ via $M_{ab}$.\footnote{Even though the equation of motion for $h$ is linear, the decomposition of $h$ based on the eigenmodes of the stationary kink does not lead to a decoupled mode-by-mode evolution.} However, a closer inspection of $S_a$ and $M_{ab}$ allows us to decouple the evolution of $G_a$ at leading order in $(\gamma v)^{-1}$.

To see this, recall that $S(x,t)$ and $\Delta W(x,t)$ at any given time have a spatial extent which scales as $\gamma^{-1}$ because of Lorentz contraction (see Fig. \ref{fig-Int123}(a) and Fig. \ref{fig-SW}). Hence
\begin{equation}
S_a(t)\sim M_{ab}(t)\sim 1/\gamma. \label{eq-sourcescaling}
\end{equation}
With these considerations in mind let us now expand $G_a$ as
\Beq
G_a(t)&=&\sum_n G_a^{(n)}(t)~.\label{eq-serie}
\Eeq
with the ansatz that $G_a^{(n)}\sim 1/(\gamma v)^{-n}$ for $t\sim\tint$. We will see that this ansatz is indeed confirmed at the end of the calculation. Using this expansion in \eqn{eq-Ga}, assuming that $S_a\sim M_{ab}\sim 1/(\gamma v)$, and collecting terms order by order in $1/(\gamma v)$ we get
\begin{eqnarray}
{G_a^{(1)}}''(t) + E_a G_a^{(1)}(t)&=&-S_a(t)~,\label{eq-1st}\\
{G_a^{(n)}}''(t) + E_a G_a^{(n)}(t) &=&-\sum_b M_{ab}(t) G_b^{(n-1)}(t)~\quad\quad\quad n\ge2.\label{eq-rec}
\end{eqnarray}
For $v\rightarrow 1$, $\gamma v\rightarrow \gamma$. Hence we have taken $\mathcal{O}[\gamma^{-1}]=\mathcal{O}[(\gamma v)^{-1}]$. As promised, $G_a^{(1)}(t)$ is sourced by the projected source term $S_a$ only; mixing via the transition matrix $M_{ab}$ only occurs at the next order and beyond. 

The solution to the above \eqn{eq-1st} and \eqn{eq-rec} is given by
\begin{eqnarray}
G_a^{(1)}(t)&=&-\int_{-\tint}^{t} d\tau \frac{\sin[\sqrt{E_a}(t-\tau)]}{\sqrt{E_a}}S_a(\tau)~,\label{eq-Ga1}\\
G_a^{(n)}(t)&=&-\sum_b\int_{-\tint}^{t} d\tau \frac{\sin[\sqrt{E_a}(t-\tau)]}{\sqrt{E_a}}M_{ab}(\tau)G_b^{(n-1)}(\tau)\quad\quad n\ge2,\label{eq-Gan}
\end{eqnarray}
where $\sin [\sqrt{E_a}(t-\tau)]/\sqrt{E_a}$ is the Green's function for the operator $\partial_t^2+E_a$. For the zero mode with $a=0$ and $E_0=0$, the above solutions take the form
\begin{eqnarray}
G_0^{(1)}(t)&=&-\int_{-\tint}^{t} d\tau(t-\tau)S_0(\tau)~,\label{eq-G01}\\
G_0^{(n)}(t)&=&-\sum_b\int_{-\tint}^{t} d\tau (t-\tau)M_{0b}(\tau)G_b^{(n-1)}(\tau)~,~n\ge2. \label{eq-sourcedG} \label{eq-G0n}
\end{eqnarray}
Based on our assumptions, the superposition of the two solitary waves is an exact solution for $t<-\tint$. Hence, in writing down the above solutions we have assumed $G^{(n)}_a(t<-\tint)=0$.

%~~~~~~~~~~~~~~~~~~~~~~~~~~figure Int 123~~~~~~~~~~~~~~~~~~~~~~~~~~~~~~~~~~~~~~~
%~~~~~~~~~~~~~~~~~~~~~~~~~~~~~~~~~~~~~~~~~~~~~~~~~~~~~~~~~~~~~~~~~~~~~~ 

Let us now concentrate on the solution for $G_a^{(1)}(t)$ for $t>\tint$:
\begin{eqnarray}
G_a^{(1)}(t>\tint)&=&A_a^{(1)}\cos\sqrt{E_a}t+B_a^{(1)}\sin\sqrt{E_a}t~,
\end{eqnarray}
while the zero mode ($E_0=0$) satisfies
\begin{eqnarray}
G_0^{(1)}(t>\tint)&=&A_0^{(1)}+B_0^{(1)}t~.
\end{eqnarray}
For $t>\tint$, the collision is by definition over and $S_a(t>\tint)=0$. $G_a^{(1)}(t>\tint)$ satisfy the ``free field" equation (\ref{eq-ga}). The non-zero coefficients $A^{(1)}_a$ and $B^{(1)}_a$ (including $A^{(1)}_0$ and $B^{(1)}_0$) are generated by the collision through $S_a(|t|<\tint)\ne0$. While the linear $t$ dependence might seem peculiar, it simply reflects the fact that the previously stationary solitary wave can be set into motion by the collision. Furthermore, we will show below that this velocity change is zero at leading order (i.e. $B_0^{(1)}=0$).

Also note that  for calculating $G_a^{(n>1)}(t)$, we require knowledge of the orthonormal basis $\{f_a(x)\}$ -- an endeavour which we will postpone to a later publication. In the next subsection, we will focus on the zero-mode and the explicit evaluation of $A_0^{(1)}$ and $B_0^{(1)}$. These coefficients are related to the phase shift and velocity change of the stationary solitary wave at leading order in $(\gamma v)^{-1}$ (see \eqn{eq-xA0shift} and \eqn{eq-vB0shift}).

%~~~~~~~~~~~~~~~~~~~~~~~~~~~~~~figure fieldplot~~~~~~~~~~~~~~~~~~~~~~~~~~~~~~~~~~~~~~
%~~~~~~~~~~~~~~~~~~~~~~~~~~~~~~~~~~~~~~~~~~~~~~~~~~~~~~~~~~~~~~~~~~~~~~
\subsection{Phase Shift and Velocity Change Calculation}
Recall that
\begin{eqnarray}
G_0^{(1)}(t)
&=&-\int_{-\tint}^{t} d\tau(t-\tau)S_0(\tau)~, \nonumber\\
&=&A_0^{(1)}+B_0^{(1)} t \quad\quad\quad\quad\quad{\rm{for}}\quad t>\tint~.
\end{eqnarray}
where $A_0^{(1)}$ and $B^{(1)}_0$ are given by
\begin{eqnarray}
A_0^{(1)}&=&\int_{-\tint}^{\tint}d\tau\, \tau S_0(\tau)~,\\
B_0^{(1)}&=&-\int_{-\tint}^{\tint}d\tau\, S_0(\tau)~.
\end{eqnarray}
where we have used $S_0(|t|>\tint)=0$ for setting the limits of integration.
The detailed calculation of $A_0^{(1)}$ and $B_0^{(1)}$ is a bit involved, however the results are exceptionally simple. At the end of the calculation, we will find
\begin{eqnarray}
A_0^{(1)}&=&\frac{M^{-1/2}}{2(\gamma v)}\int_{0}^{\Delta\phi}\int_{0}^{\Delta\phi}d\phi_1d\phi_2\left[\frac{V(\phi_1+\phi_2)-V(\phi_1)-V(\phi_2)}{\sqrt{V(\phi_1)V(\phi_2)}}\right]+\mathcal{O}[(\gamma v)^{-2}]~,\\
B_0^{(1)}&=&0~.
\end{eqnarray}

%~~~~~~~~~~~~~~~~~subsub Velocity Change ~~~~~~~~~~~~~~~~~~~~~~~~~~~~
\subsubsection{Velocity Change}
Let us calculate $B_0^{(1)}$ as follows:
\begin{eqnarray}
B_0^{(1)}&=&-\int_{-\tint}^{\tint}d\tau\, S_0(\tau)~, \nonumber\\
&=&-\int_{\Aint}d\tau d\chi\, f_0(\chi)S(\chi,\tau)~, \nonumber\\
&=&M^{-1/2}\int_{\Aint}d\tau d\chi\, \phi_K'(\chi)\left\{V'[\phi_K(\chi)]+V'[\phi_K[\gamma(\chi+v \tau)]\right.\nonumber\\ 
& &\quad\quad\quad\quad\quad
-\left.V'[\phi_K(\chi)+\phi_K[\gamma(\chi+v\tau)]]\right\}~.
\end{eqnarray}
In the second line we used the the definition of the projected source $S_0(t)=\int f_0(x)S(x,t)$ and fact that $S(x,t)=0$ unless we are within the interaction area $\Aint$ shown in Fig. \ref{fig-Int123}(a).  In the third line we used the definition of $S(x,t)$ and $f_0(x)$ in equations (\ref{eq-source}) and (\ref{eq-f0def}). The integrations simplify considerably if we make the following co-ordinate transformation (see Fig. \ref{fig-Int123}(a) and (b))
\begin{eqnarray}
\left(\chi,\tau\right)&=&\left(q,\frac{p}{\gamma v}-\frac{q}{v}\right)~.
\end{eqnarray}
Under this transformation, the area element and integration region transform as 
\begin{eqnarray}
&d\chi d\tau=(\gamma v)^{-1}dpdq~,\\
&\Aint=\Aint^{pq}~.
\end{eqnarray}
Putting everything together in the expression for $B_0^{(1)}$ we get
\begin{eqnarray}
B_0^{(1)}
&=&\frac{M^{-1/2}}{(\gamma v)}\int_{\Aint^{pq}}dp dq\, \phi_K'(q)\left\{V'[\phi_K(q)]+V'[\phi_K(p)]-
V'[\phi_K(q)+\phi_K(p)]\right\}~.\nonumber \\ \label{eq-B0}
\end{eqnarray}
Let us calculate each of these three terms separately. Each will be zero. The key step in the manipulations is the following. Based on our localization assumption stated in equation (\ref{eq-prac}), we assume that $\phi=0$ for $x\le-L/2$ and $\phi=\Delta\phi$ for $x\ge L/2$.  With that in mind let us look at the first term in \eqn{eq-B0}
\begin{eqnarray}
\int_{\Aint^{pq}}dpdq\,\phi_K'(q)V'[\phi_K(q)]~ 
&=&\int_{-L/2}^{L/2} dp\,\int_{0}^{\Delta\phi}d\phi V'(\phi)~, \nn
&=&\int_{-L/2}^{L/2} dp\,\left[V(0)-V(\Delta \phi)\right]=0~,
\end{eqnarray}
since $V(0)=V(\Delta\phi)=0$.
For the second term  in \eqn{eq-B0}, we have
\begin{eqnarray}
\int_{\Aint^{pq}}dpdq\,\phi_K'(q)V'[\phi_K(p)]
&=&\int_{\Aint^{pq}} dpdq\,\phi_K'(q) \phi_K''(p)~, \nonumber\\
&=&\int_{-L/2}^{L/2} \, dq\phi_K'(q) \left[\phi_K'(L/2)-\phi_K'(-L/2)\right]~,\nonumber\\ 
&=&0~.
\end{eqnarray}
Above, we used $\phi_K''(p)=V'[\phi_K(p)]$ and $\phi_K'(L/2)=\phi_K'(-L/2)=0$. Finally, for the third term in \eqn{eq-B0} we have
\begin{eqnarray}
\nonumber
\int_{\Aint^{pq}}dpdq\,\phi_K'(q)V'[\phi_K(q)+\phi_K(p)]
&=&\int_{-L/2}^{L/2} dp \int_{0}^{\Delta\phi}\,d\phi V'[\phi+\phi_K(p)]~,\\ \nonumber
&=&\int_{-L/2}^{L/2} dp \left(V[\Delta\phi+\phi_K(p)]-V[\phi_K(p)]\right)~,\\
&=&0.
\end{eqnarray}
where we used $V(\phi+\Delta\phi)=V(\phi)$. 

In summary, we have just shown that
\begin{eqnarray}
B_0^{(1)}&=&0~,\\
\Delta v
&=&-\frac{1}{\sqrt{M}}B_0^{(1)}+\mathcal{O}[(\gamma v)^{-2}]~,\nonumber\\
&=&0+\mathcal{O}[(\gamma v)^{-2}]~.
\end{eqnarray}
%~~~~~~~~~~~~~~~~~subsub Phase Shift ~~~~~~~~~~~~~~~~~~~~~~~~~~~~
\subsubsection{Phase Shift}
Let us now turn our attention to $A_0^{(1)}$:
\begin{eqnarray}
A_0^{(1)}
&=&\int_{-\tint}^{\tint}d\tau\, \tau S_0(\tau)~,  \nonumber\\
&=-&M^{-1/2}\int_{\Aint}d\tau d\chi\,\tau \phi_K'(\chi)\left\{V'[\phi_K(\chi)]+V'[\phi_K[\gamma(\chi+v \tau)]\right. \nonumber\\ 
& &\qquad-\left.V'[\phi_K(\chi)+\phi_K[\gamma(\chi+v\tau)]]\right\}~, \nonumber\\
&=&-\frac{M^{-1/2}}{(\gamma v)}\int_{\Aint^{pq}}dp dq\, \left(\frac{p}{\gamma v}-\frac{q}{v}\right)\phi_K'(q)\left\{V'[\phi_K(q)]+V'[\phi_K(p)]\right. \nonumber\\
& &\qquad-\left.V'[\phi_K(q)+\phi_K(p)]\right\}~,\nonumber\\ 
&=&\frac{M^{-1/2}}{(\gamma v)}\frac{1}{v}\int_{\Aint^{pq}}dp dq\, q\phi_K'(q)\left\{V'[\phi_K(q)]+V'[\phi_K(p)]-
V'[\phi_K(q)+\phi_K(p)]\right\}\nonumber\\ 
& &\qquad-\frac{M^{-1/2}}{(\gamma v)^2}\int_{\Aint^{pq}}dp dq\, p\phi_K'(q)\left\{V'[\phi_K(q)]+V'[\phi_K(p)]-
V'[\phi_K(q)+\phi_K(p)]\right\}~.\nonumber\\
\label{eq-A0}
\end{eqnarray}
Notice the extra $\tau$ factor in the integrand of $A_0^{(1)}$ becomes $\tau=p/(\gamma v)-q/v$. As in the case of $B_0^{(1)}$, we will integrate each of the above terms separately.

For the first term in equation (\ref{eq-A0}), the integral is 
\begin{eqnarray}
\nonumber
\int_{\Aint^{pq}}dp dq\, q\phi_K'(q)V'[\phi_K(q)]
&=&\int_{-L/2}^{L/2} dp \int_{0}^{\Delta\phi}d\phi_1 q(\phi_1)V'(\phi_1)~,\\ \nonumber
&=-&\int_{-L/2}^{L/2} dp\int_{0}^{\Delta\phi}d\phi_1 q'(\phi_1)V(\phi_1)~,\\ \nonumber
&=-&\int_{0}^{\Delta\phi} d\phi_2 p'(\phi_2)\int_{0}^{\Delta\phi}d\phi_1 q'(\phi_1)V(\phi_1)~,\\ 
&=-&\frac{1}{2}\int_{0}^{\Delta\phi}\int_{0}^{\Delta\phi}\frac{d\phi_1 d\phi_2 }{\sqrt{V(\phi_1)V(\phi_2)}}V(\phi_1)~.
\end{eqnarray}
In the second line we integrated by parts with boundary terms giving no contribution. The third and fourth line converts the co-ordinate space variables to the field space variables. In the fifth line we used $\phi_K'(x)=\sqrt{2V(\phi_K(x))}$ which can be obtained by integrating the equation of motion, $\phi_K''(x)=V'[\phi_K(x)]$ by parts. Note that $p(\phi)$ and $q(\phi)$ are invertible in the range of interest. 

The second term in equation (\ref{eq-A0}) can be converted to a boundary term of the $p$ co-ordinate and evaluates to zero.
\begin{eqnarray}
\nonumber
\int_{\Aint^{pq}}dpdq\,q\phi_K'(q)V'[\phi_K(p)]
&=&\int_{\Aint^{pq}} dpdq q\,\phi_K'(q) \phi_K''(p)~,\\ \nonumber
&=&\int_{-L/2}^{L/2} \, dq q \phi_K'(q) \left[\phi_K'(L/2)-\phi_K'(-L/2)\right]~,
\nonumber \\ 
&=&0.
\end{eqnarray}
For the third term in equation (\ref{eq-A0}) we have
\begin{eqnarray}
 \nonumber
& &\int_{\Aint^{pq}}dp dq\, q\phi_K'(q)V'[\phi_K(q)+\phi_K(p)]~\\ \nn
&=&\int_{0}^{\Delta\phi}\frac{d\phi_2 }{\sqrt{2V(\phi_2)}}\int_{0}^{\Delta\phi}d\phi_1q(\phi_1)V'[\phi_1+\phi_2]~,\\ \nonumber
&=&\int_{0}^{\Delta\phi}\frac{d\phi_2 }{\sqrt{2V(\phi_2)}}\left[q(\phi_1)V(\phi_1+\phi_2)\Bigg|_{0}^{\Delta\phi}-\int_{0}^{\Delta\phi}d\phi_1q'(\phi_1)V(\phi_1+\phi_2)\right]~,\\ \nonumber
&=&\int_{0}^{\Delta\phi}\frac{d\phi_2 }{\sqrt{2V(\phi_2)}}\left[\int_{0}^{\Delta\phi}\frac{d\phi_1}{\sqrt{2V(\phi_1)}}V(\phi_2)-\int_{0}^{\Delta\phi}\frac{d\phi_1}{\sqrt{2V(\phi_1)}}V(\phi_1+\phi_2)\right]~,\\ 
&=&\frac{1}{2}\int_{0}^{\Delta\phi}\int_{0}^{\Delta\phi}\frac{d\phi_2 d\phi_2}{\sqrt{V(\phi_1)V(\phi_2)}}\left[V(\phi_2)-V(\phi_1+\phi_2)\right]~.
\end{eqnarray}
In line three above, we used 
\begin{eqnarray}
\nonumber
q(\phi_1)V(\phi_1+\phi_2)\Bigg|_{\phi_1=0}^{\phi_1=\Delta\phi}
&=&q(\Delta\phi)V(\Delta\phi+\phi_2)-q(0)V(0+\phi_2)~,\\ \nonumber
&=&\left[q(\Delta\phi)-q(0)\right]V(\phi_2)~,\\ 
&=&\int_{0}^{\Delta\phi}\frac{d\phi_1}{\sqrt{2V(\phi_1)}}V(\phi_2)~.
\end{eqnarray}
With similar manipulations, one can show that the fourth, and sixth term of equation (\ref{eq-A0}) are zero whereas the fifth evaluates to $-M^{-1/2}\Delta \phi^2/(\gamma v)^2$.
Combining all of these results, we can finally write down a closed form expression for $A_0^{(1)}$ in equation (\ref{eq-A0}):
\begin{eqnarray}
A_0^{(1)}&=&\frac{M^{-1/2}}{2(\gamma v)}\left\{\frac{1}{v}\int_{0}^{\Delta\phi}\int_{0}^{\Delta\phi}d\phi_1d\phi_2\left[\frac{V(\phi_1+\phi_2)-V(\phi_1)-V(\phi_2)}{\sqrt{V(\phi_1)V(\phi_2)}}\right] -\frac{2}{(\gamma v)}\Delta \phi^2\right\}~,\nonumber\\
&=&\frac{M^{-1/2}}{2(\gamma v)}\int_{0}^{\Delta\phi}\int_{0}^{\Delta\phi}d\phi_1d\phi_2\left[\frac{V(\phi_1+\phi_2)-V(\phi_1)-V(\phi_2)}{\sqrt{V(\phi_1)V(\phi_2)}}\right]+\mathcal{O}[(\gamma v)^{-2}]~.
\end{eqnarray}
where in the last line we have kept the leading order term in $(\gamma v)^{-1}$. With this $A_0^{(1)}$, the phase shift is
\begin{eqnarray}
\label{eq-PhaseShift} 
\Delta x
&=&-\frac{A_0^{(1)}}{\sqrt{M}}+\mathcal{O}[(\gamma v)^{-2}]~,\\
&=&\frac{1}{2(\gamma v)M}\int_{0}^{\Delta\phi}\int_{0}^{\Delta\phi}d\phi_1d\phi_2\left[\frac{V(\phi_1)+V(\phi_2)-V(\phi_1+\phi_2)}{\sqrt{V(\phi_1)V(\phi_2)}}\right]+\mathcal{O}[(\gamma v)^{-2}]~.\nonumber
\end{eqnarray}
Note that in the second line above, $\mathcal{O}[(\gamma v)^{-2}]$ contains higher order corrections to $A_0^{(1)}$ calculated in this subsection, as well as corrections from $n>1$ terms in equation (\ref{eq-G0n}).

Being able to write the kink-kink interaction as a simple integral of $V$ in field space is a very powerful tool.  For example, since the field profile of an antikink is given by $\phi_K(-x)$, it is straightforward to verify that equation (\ref{eq-PhaseShift}) also gives the phase shift between a pair of kink-antikink.  We can hence conclude that the leading order result of a kink-kink collision and a kink-antikink collision are identical. The sign of the phase shift depends on the details of the potential. As we will show in the next section, it is possible to get positive and negative phase shifts corresponding to an attractive and repulsive interaction respectively.

In the next section we check our results for the phase shift and velocity change for a number of examples. For the Sine-Gordon case, these quantities are compared to the exact results. For other cases, we compare our answers to those obtained by full numerical integrations of the equations of motion. We will find that our order by order results agree exceptionally well with both exact results (when available) and numerical simulations of collisions.

%~~~~~~~~~~~~~~~~ Examples ~~~~~~~~~~~~~~~~~~~~~~~~~~~~~~~~
%%%%%%%%%%%%%%%%%%%%%%%%%%%%%%%%%%%%%%
\section{Examples}
\label{sec-ex}

%~~~~~~~~~~~~~~~~ sub Sine-Gordon~~~~~~~~~~~~~~~~~~~~~~~~~~~~~~~~
\subsection{Sine-Gordon}

Consider the normalized Sine-Gordon potential,
\begin{equation}
V(\phi) = 1-\cos\phi~.
\label{eq-SG}
\end{equation}
Let us calculate the leading order phase shift in this model based on our result (\ref{eq-PhaseShift}). For the Sine-Gordon case, the kink solution is given by $\phi_K(x)= 4\tan^{-1}[e^x]$. From equation (\ref{eq-mass}), we get $M=8$ and using (\ref{eq-PhaseShift}), the phase shift is
\Beq
\Delta x=\frac{2}{(\gamma v)}+\mathcal{O}[(\gamma v)^{-2}]~.
\Eeq
In the Sine-Gordon case, the two kink solution can be obtained analytically, based on which one can calculate the phase shift exactly\cite{McLAlw73,ScoChu73,SWRev80}.
\Beq
\Delta x= \ln \left(\frac{\gamma+1}{\gamma-1}\right)=\frac{2}{(\gamma v)}+\mathcal{O}[(\gamma v)^{-3}]~.
\Eeq
Thus, the phase shift calculated based on equation (\ref{eq-PhaseShift}) at leading order agrees exactly with the phase shift (again at leading order) based on the exact Sine-Gordon kink-kink (or kink-antikink or antikink-antikink) solution. Also note that for the Sine-Gordon case, the exact solution shows that there is no velocity change due to the collision. This is again consistent with our leading order result $\Delta v=0+\mathcal{O}[(\gamma v)^{-2}]$. \footnote{In \cite{Mal85a,Mal85b} small deformation from the Sine-Gordon case were studied, and they reach the same conclusion that $\Delta v \sim (\gamma v)^{-2}$.}

%~~~~~~~~~~~~~~~~ sub Away from SG~~~~~~~~~~~~~~~~~~~~~~~~~~~~~~~~

\subsection{Away from Sine-Gordon}
We now consider models for which analytic solutions are not known. For concreteness we consider models of the form
\Beq
V(\phi)=(1-\cos\phi)(1-\alpha \sin^2\phi)~,
\label{eq-PotEps}
\Eeq
where $-1<\alpha<1$. Note that these potentials are periodic with a period $\Delta\phi=2\pi$. Importantly, these models are not necessarily small deformations of the Sine-Gordon case ($\alpha=0$). %Also note that while the above potential ossimilar results were obtained when the potential in equation \ref{eq-PotEps} was odd rather than even 

We numerically simulate the collision of two kinks in this model. Initially, one of the kinks is stationary and another is moving towards the stationary one from the positive to negative direction. The stationary kink profile is obtained numerically by a relaxation technique\footnote{We introduce an additional friction term in the equations of motion and allow the solution to relax to a time independent solution. This time independent solution is the stationary kink profile we want.}. A Lorentz transformation of the stationary kink profile is then used to obtain the profile of the incoming kink. We take a superposition of the two profiles to obtain initial conditions for the collision. The initial conditions used are shown in Figure \ref{fig-FieldPlot}(top). The equation of motion for the field is evolved using a 4th order Runge Kutta method (with rigid boundary conditions). 

After the collision, we calculate the phase and velocity shift of the stationary solitary wave from the numerically evolved $\phi(x,t)$  as follows. First, for $\tint<t\lesssim 10\tint$ we carry out the following projection: 
\Beq
I(t)
&=&\int_{-L}^{L} dxf_0(x)\left[\phi(x,t)-\phi_K(x)-\phi_K(\gamma(x+vt)\right]~,\\
&=&\int_{-L}^{L} dxf_0(x)\left[\phi(x,t)-\phi_K(x)-2\pi\right]~.
\Eeq
Note the $2\pi$ arises from the asymptotic value of the second kink after it has moved away from the stationary kink. We then fit this numerically calculated $I(t)$ to a line $\tilde{A}_0+\tilde{B}_0 t$, and determine the coefficients $\tilde{A}_0$ and $\tilde{B}_0$. These coefficients are directly related to the phase and velocity shift of the stationary kink. To understand why we fit $I(t)$ to a straight line, and why these numerically calculated coefficients provide a measure of the phase and velocity shift, let us express  $I(t)$ in terms of our eigenmodes $f_a(x)$ and their coefficients $G_a(t)$. For $t>\tint$:
\Beq
\nonumber
 I(t)&=&\int_{-L}^{L}dx f_0(x)h(x,t)~,\\ \nonumber
 &=&\int_{-L}^{L}dxf_0(x)\sum_a G_a(t) f_a(x)~,\\ \nonumber
 &=&G_0(t)~,\\ \nonumber
 &=&G_0^{(1)}(t)+\sum_{n> 1} G^{(n)}_0(t)~,\\ 
 &=&A_0^{(1)}+B_0^{(1)} t+\sum_{n > 1} G^{(n)}_0(t)~.
 \Eeq
where in the the second line we expanded $h(x,t)$ in terms of eigenmodes, in the third line we used the orthonormality of $f_a(x)$\footnote{Note that for orthonormality we need $L\rightarrow \infty$, however in practice the localization of $f_0(x)$ to $\sim L$ allows us to cut off the integral at a finite $L$.}, in the fourth line we expanded $G_0(t)$ as a series and in the fifth line we explicitly write down the leading order term in the form of its eigenmode (\ref{eq-g0def}). As we have shown earlier, $G_0^{(n)}(t\sim \tint)\sim 1/(\gamma v)^{-n}$ with $(\gamma v)\gg 1$. Hence, when we fit the numerically calculated $I(t)$ to a straight line $\tilde{A}_0+\tilde{B}_0 t$, we are estimating the coefficients $A_0^{(1)}$ and $B_0^{(1)}$. Once these co-efficients have been estimated numerically, it is easy to find the phase shift and the velocity change of the stationary kink via equations (\ref{eq-xA0shift}) and (\ref{eq-vB0shift}).%These coefficients are then directly related to the phase and velocity shifts via equations (\ref{}) and (\ref{})
 
 %One needs to be large enough to have support from $f_0$  and also assumed that $L of $f_a(x)$ (\ref{eq-ortho}) we have $I(t)=G_0(t)=A_0+B_0t$ to leading order in $(\gamma v)^{-1}$. %Fitting the numerical calculated $I(t)$ to a line $A_0^N+B_0^Nt$, we can get $A_0^N$ and $B_0^N$ which are then used to calculate the numerical phase and velocity shifts.
%~~~~~~~~~~~~~~~~~~~fig PhaseShiftG~~~~~~~~~~~~~~~~~~~~~~~~~~~~~~~~~~~~~~~~~~
\begin{figure}
\begin{center}
\includegraphics[width=14cm]{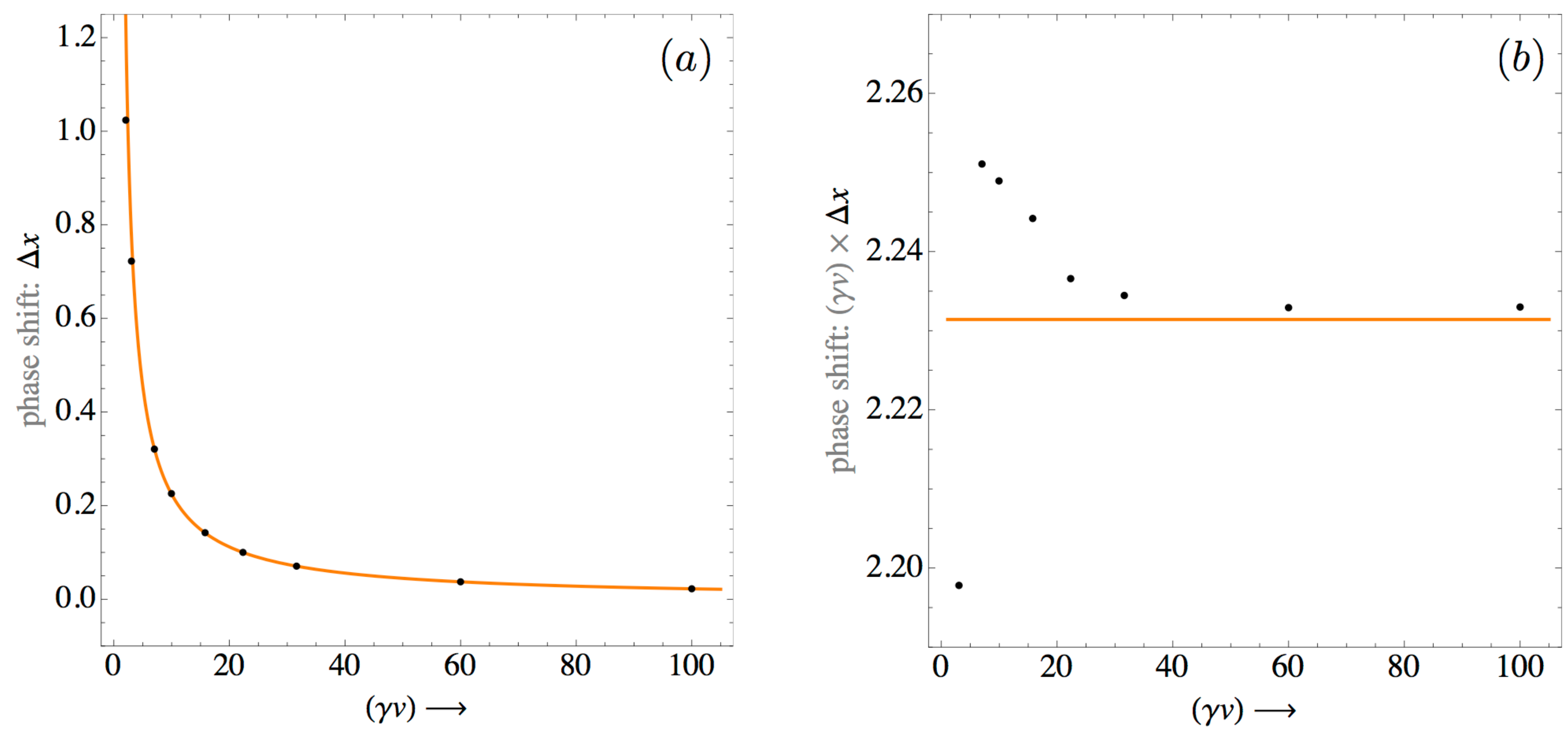}
\caption{In (a), we plot the numerically calculated phase shift undergone by a stationary kink colliding with an incoming kink as a function of $(\gamma v)$. For this plot, the scalar field potential $V(\phi)=(1-\cos \phi)(1-0.5\sin^2\phi)$. The orange curve is the theoretical prediction at leading order in $1/(\gamma v)$ and the black dots are the simulation results. They are in a excellent agreement. In (b), we multiply the phase shift by $(\gamma v)$ to show how the numerical and analytic calculations approach each other as $(\gamma v)$ increases.
\label{fig-PhaseShiftG}}
\end{center}
\end{figure}
%~~~~~~~~~~~~~~~~~~~~~~~~~~~~~~~~~~~~~~~~~~~~~~~~~~~~~~~~~~~~~~~~~~~~~~
\begin{figure}
\begin{center}
\includegraphics[width=7cm]{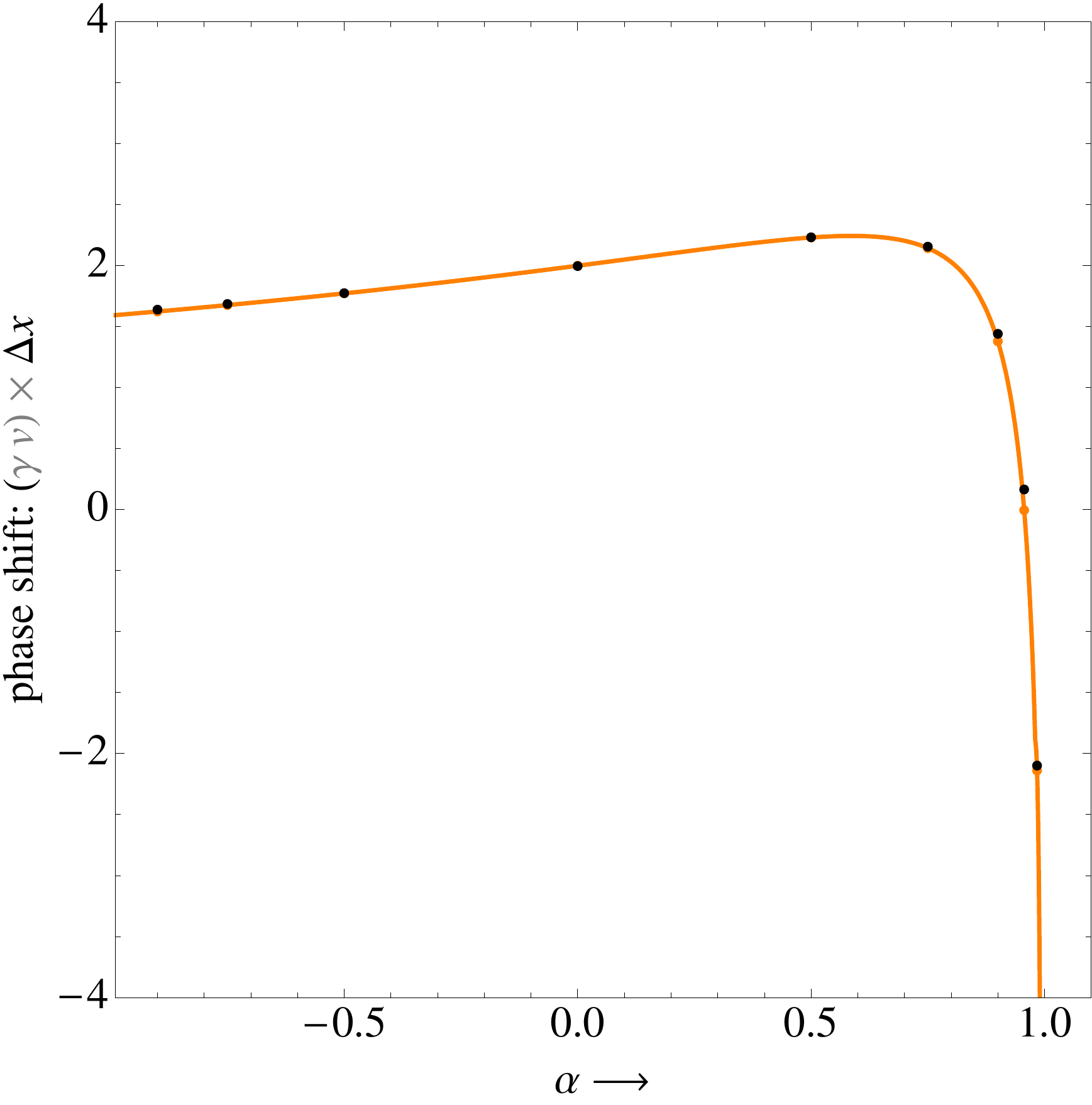}
\caption{The phase shift (multiplied by $(\gamma v)$) undergone by a stationary kink colliding with an incoming kink as a function of the $\alpha$ parameter in the potential $V(\phi)=(1-\cos \phi)(1-\alpha\sin^2\phi)$. For this plot $(\gamma v)=100$. The orange curve (and orange dots) is the theoretical prediction at leading order in $(\gamma v)^{-1}$ and the black dots are the simulation results. The agreement between the two is excellent. %At lower $(\gamma v)$ the differences between the analytic and numerical results start becoming evident near $\alpha \rightarrow 1$, the domain where the phase shift starts changing its sign.
}
\label{fig-PhaseShiftE}
\end{center}
\end{figure}
%~~~~~~~~~~~~~~~~~~~~~~~~~~~~~~~~~~~~~~~~~~~~~~~~~~~~~~~~~~~~~~~~~~~~~
We carried out a large number of high resolution simulations of the collisions, varying both the velocity of the incoming kink $v$ and parameter $\alpha$ in the potential (\ref{eq-PotEps}). Using the projection $I(t)$ discussed above, we calculated the phase shift and the velocity change of the stationary kink as a function $(\gamma v)$ and $\alpha$. We summarize our results below.

Fig. \ref{fig-PhaseShiftG} shows the comparison of the phase shift calculated based on equation (\ref{eq-PhaseShift}) and that from the numerical simulations. For this figure, we used $\alpha=0.5$ and varied the speed $v$ of the incoming kink. Note that the $(\gamma v)$ dependence is wonderfully captured by our leading order result, even when $(\gamma v)\approx 3$. As expected, the difference between the numerically calculated phase shift and the one based on equation (\ref{eq-PhaseShift}) diminishes as $(\gamma v)$ increases.

Next we check the $\alpha$ dependence of the phase shift at a fixed $\gamma v=100$. As seen in Fig. \ref{fig-PhaseShiftE}, the phase shift calculated from the numerical simulations matches well with the leading order result. The reason we fixed $(\gamma v)=100$ is interesting in its own right. When computing $I(t)$ numerically, we found that when $\alpha$ was not too close to $1$, the deviation of $I(t)$ from a straight line fit (used to determine the phase shift) was oscillatory and quite small, scaling with inverse powers of $(\gamma v)$. However, for the same $(\gamma v)$, when $\alpha \rightarrow 1$, the deviations from a straight line fit were quite large. We found that for $\alpha>0.9$, one has to go to sufficiently high $(\gamma v)$ to see a good match between the analytically and numerically calculated phase shifts. It is also worth noting that this is the region where the interaction starts becoming repulsive ($\Delta x$ changes sign). Thus, the $(\gamma v)$ at which our leading order results provide a good approximation can vary with model parameters. In the future, we will explore the relationship between this rate of convergence and the functional form of the potential. 

Finally, while we do not show the result here, we also checked that the velocity shift $\Delta v=0+\mathcal{O}[(\gamma v)^{-2}]$. 

In summary, the calculated phase and velocity shift based on our $(\gamma v)^{-1}$ expansion are in excellent agreement with the results from numerical simulations of ultra relativistic kinks.%\footnote{The numerically calculated $\Delta x\propto 1/\gamma$ to a better accuracy than to $1/(\gamma v)$. The corrections to the leading order $1/\gamma$ scaling, also seem to follow a power law (which is not the case for $1/(\gamma v)$).}
%~~~~~~~~~~~~~~~~ Discussion  ~~~~~~~~~~~~~~~~~~~~~~~~~~~~~~~~~~~~~~~
%%%%%%%%%%%%%%%%%%%%%%%%%%%%%%%%%%%%%%%%%%%%%%%%%%%%%%%%%%%%
\section{Discussion}
\label{sec-dis}
In this paper, we established a general framework to calculate the outcome of ultra- relativistic collisions between solitary waves in relativistic scalar field theories. We showed that the colliding solitary waves pass through each other, and the perturbations to this free passage behavior are small due to the suppression of the space-time area of interaction $\Aint\propto 1/(\gamma v)$ where the two solitary waves overlap. We present an order by order prescription to calculate the full result of the collision.

We considered collisions of localized, quasi-stable, scalar-field solitary waves in periodic potentials or potentials with a single minimum. In our set-up, an ultra relativistic solitary wave ($\gamma v\gg 1$) collides with a stationary solitary wave. We showed that for linearized perturbations, the stationary solitary wave's perturbations can be organized as a power series in $1/(\gamma v)$. For small amplitude perturbations, the corrections from nonlinear effects can also be expressed as a power series, as shown in Appendix~\ref{sec-nonlinear}. In such cases, there exists a $(\gamma v)$ high enough that the full result is under analytical control, and well approximated by the leading order effects. %In certain cases, it is also possible to express the leading order effects in closed forms. 

We applied our formalism to a specific example: $(1+1)$ dimensional kinks with periodic potentials. We calculated two leading order effects with important physical meanings: phase shift and velocity change. We showed that the leading order results can be expressed in closed forms in terms of the potential. In other words we can know the leading order result before they collide---making analytical predictions which can be checked with experiments (simulations). 

We showed that the leading order phase shift for ultrarelativistic collisions is independent of whether the collision is between kinks  or between a kink and an antikink. Although there is no direct contradiction, we note that the long-range interaction between kinks is repulsive, whereas between a kink and antikink is attractive \cite{Manton:1978gf}. We were also able to construct examples with a zero or a negative phase shift at leading order.

Collisions of solitary waves have been investigated analytically before in a very special and limited subset---integrable (e.g Sine-Gordon) or approximately integrable cases. Our results agree with these cases. More importantly, we showed that for a potential that was arbitrarily far from being integrable, our prediction still agreed extremely well with numerical simulations. 

In summary, understanding soliton interactions has been an active area of research for more than 50 years. Many interesting physical phenomena involve solitons such as fluxons in Josephson junctions \cite{McLSco78}, non-linear optical solitons \cite{BulCau80}, reheating after inflation \cite{Amin:2011hj} and domain wall collisions in cosmology \cite{EasGib09}. Apart from numerical techniques, there are two standard approaches. One is to model them as being perturbatively close to the integrable Sine-Gordon system. Another approach is to carry out a dynamical systems analysis of the collective-coordinates ordinary differential equations \cite{ZhuHab08, MalOptRev, PhysRevLett.98.104103}. Here, we demonstrated a novel third method -- a kinematics based scattering theory at relativistic velocities. Our method works well for collisions at ultrarelativistic velocities, which is exactly when numerical techniques become inefficient. For these collisions, we do not rely on a small deformation from Sine-Gordon, thus our analytical framework is applicable to a wider range of phenomena.

\subsection*{Future Directions}
%After many years of simulations and/or hunting for isolated integrable systems, we believe that our work provides a stepping stone towards a general analytical method for understanding solitary wave collisions. While we have provided a general framework, a number of avenues still remain unexplored. 
%Below, we provide a list of some future directions.
\begin{itemize}

\item Testing the framework with examples:\\
Perhaps the most natural next step is to test our framework with different examples. Collisions of oscillons, $Q$-balls and bubbles in $1+1$ and higher dimensions can be simulated and compared with the $1/(\gamma v)$ behavior (of the leading order effects) predicted by our framework. It would be also be interesting to see if collisions of localized objects composed of multiple fields still respect our framework.

\item The localization condition: \\
We have focused our attention on solitary waves with exponentially suppressed tails. That is clearly sufficient but not necessary. In particular, the earliest observation of free passage occurs in strings (vortices)\cite{She87}. For them the tails are not only power-law, but also lead to infinite integrated energy. However it still has a clean relativistic collision and it might be described by a method similar to our framework. It will be interesting to figure out the most general class of objects that their relativistic collisions allow full analytical descriptions.

\item Non-Lorentz invariant theories:\\
We have focused our attention on Lorentz invariant scalar field theories. However, certain classic systems such shallow water waves described by the KdV equation \cite{KdV} are not Lorentz invariant. Nevertheless, they contain two solitary wave solutions where the phase shift does decrease with velocity in a manner reminiscent of our results in this paper. It would be interesting to see whether such systems can still be described within our framework.
\item Higher order effects: \\
We provided the recursive equation to calculate higher order effects in the linearized theory. It is already tedious to do a calculation beyond the leading order with them. Going beyond the linearized equations makes it even more so. Although still doable, the required computation resources may exceed a direct simulation. However, in this paper we na\"ively wrote down all terms without considering symmetries which could have simplified our analysis. This is akin to drawing all Feynman diagrams without recognizing that some (maybe the majority) of them can cancel with others. To investigate such cancellations, in Appendix \ref{sec-energy} we wrote down energy conservation equations order by order. These equations play the role of the optical theorem in perturbation theory. In Appendix~\ref{sec-energy} we show that these equations are already quite powerful at leading order. They allow us to conclude that the velocity change is zero at leading order, without any detailed calculations. It is possible that similar techniques can be used to simplify the expressions for higher order effects.
\item An inverse search for integrable systems: \\
One property of integrable systems is the lack of velocity change after collisions (to all orders). Since we have an analytical expression for the velocity change, setting it to zero order by order in principle provides the set of analytical conditions for integrable systems.\footnote{We thank Adam Brown for suggesting this possibility.} This of course relies on the previous point that the full recursive series needs to be simplified to make the condition useful.
\item Gravitational Effects:\\
We have completely ignored gravity in our framework. In \cite{JohYan10}, the authors explore the gravitational effects in bubble collisions (in the context of classical transitions). Gravity can have dramatic effects in certain ultrarelativistic collisions. It has been shown by \cite{ChoPre09} that one can form black-holes by colliding ultrarelativistic solitary waves, which of course cannot be seen in our framework. It would be interesting to see whether one can appropriately incorporate gravity into our framework. 
\end{itemize}
%We hope that these ideas will be pursued further in future studies.

\section*{Acknowledgemements}
MA is supported by a Kavli Fellowship. ISY is supported by the research program of the Foundation for Fundamental Research on Matter (FOM), which is part of the Netherlands Organization for Scientific Research (NWO). MA and EAL also thank the Department of Mathematics and Theoretical Physics (DAMTP) at Cambridge where part of this work was done. We thank Daniel Baumann, Roger Blandford, Adam Brown, Cliff Burgess, George Efstathiou, Tom Giblin, Lam Hui, Nick Manton, Theodorus Nieuwenhuizen, Boris Malomed, Ignacy Sawicki, Bob Wagoner and Erick Weinberg for useful conversations. We especially thank Adam Brown, Tom Giblin and Erick Weinberg for their insightful comments and suggestions for improvement of this manuscript. MA thanks the organizers of the Primordial Cosmology workshop at KITP, Santa Barbara, 2013. EAL thanks the organizers of Peyresq Physics 18, 2013, and acknowledges the support of OLAM, Association pour la Recherche Fondamentale, Bruxelles and an FQXi minigrant for work on ``Cosmological Bubble Collisions". Some of the high resolution numerical simulations were performed on the COSMOS supercomputer, part of the DiRAC HPC, a facility which is funded by STFC and BIS.
%%%%%%%%%%%%%%%%%%%%%%%%%%%%%%%%%%%%%%%%%%%%
\bibliography{all}
\bibliographystyle{utcaps}

%%%%%%%%%%%%%%%%%%%%%%%%%%%%%%%%%%%%%%%%%%%%%
\appendix

\section{Post-collision evolution}
\label{sec-LongTime}

In the main body of the text we were able to show that soon after the collision $t\sim \tint$, the perturbations of a stationary soliton generated by the collision can be organized as a convergent series with the ratio of consecutive terms scaling as $(\gamma v)^{-1}$. However, we alluded to the fact that this scaling might be broken when evaluating the perturbations long after the collision. In this appendix we discuss this issue in detail and provide a prescription to calculate the perturbation for all time. 

To obtain the result of solitary wave collisions valid for all time, we need to carry out two parallel calculations, one in the rest frame of each solitary wave. %In a single copy of such calculation, we will first assume that the changes are small. This provides a recursive way to calculate it, and then we see that it is indeed small and our assumption is self-consistent. 
The full result of the collision is then an appropriate combination from the two parallel calculations.  As we will see, the perturbations induced by the collision are small. Hence this combination is no more than a linear superposition of the the perturbations from both calculations (in most, though not all of the space-time regions of interest).  

We suggest that the reader refer to Sec. \ref{sec-GPertExp} for definitions and some background for what is discussed below. We begin by writing the solution to the linearized equation of motion \eqn{eq-generalEOM}as a series $h(\bx,t)=\sum_{n}h^{(n)}(\bx,t)$.  Each term is then formally given by:
\Beq
h^{(1)}(\bx,t)&=&\int dt'd\bx'\mG(\bx,t;\bx',t')S(\bx',t')~, \\
h^{(n>1)}(\bx,t)&=&\int dt'd\bx\mG(\bx,t;\bx',t')\Delta W(\bx',t')h^{(n-1)}~.\label{eq-hSourceA}
\Eeq
where $\mG=(\Box-W_0)^{-1}$. If evaluated at $t\sim \tint$ and $|x|<L_A/2$, we argued in Sec. \ref{sec-GPertExp} that $h^{(n)}\propto \Aint h^{(n-1)}$ where the overlap area $\Aint\propto 1/(\gamma v)$. However, for $t\gg \tint$ this need not be the case. This is primarily because $\Delta W(\bx,t\gg \tint)\ne0$. While $S(\bx,t)$ effectively shuts off for $t>\tint$, the $\Delta W(\bx,t)$ term does not. Physically, at late times, the integral over $\Delta W h^{(n-1)}$ is related to how perturbations from the outgoing kink influence perturbations on the stationary one. %In principle, this can lead to $h^{(n>1)}$ growing with time, invalidating the heirarchy of terms $h^{(n)}\sim (\gamma v)^{-1}h^{(n-1)}$ at late times. 

Instead of the general hierarchy in \eqn{eq-hSourceA}, a better approach is to first split the perturbation $h$ into two parts 
\Beq
h(\bx,t)\approx h_A(\bx,t)+h_B(\bx,t)~,
\Eeq
where \begin{enumerate} 
\item $h_A(\bx,t)$ = perturbations on $\phi_A$ generated by the collision. $h_A$ includes all effects generated from the interaction area $\Aint$, including localized perturbations as well as outgoing radiation on and from $\phi_A$. 
\item $h_B(\bx,t)$ = incoming radiation generated (or reflected) from $\phi_B$. More precisely, $h_B$ includes possible contributions to $h$ at the point $(\bx,t)$ from the dashed orange box in Fig. \ref{fig-Int}(b). Note that the dashed orange box represents the relevant space-time occupied by $\phi_B$ beyond $\Aint$. 
\end{enumerate}

The calculation of $h_A$ is surprisingly simple. It is just the calculation of $h$ in Sec. \ref{sec-GPertExp}, restricting the integration range to $\Aint$, but allowing it to be generally valid even for $t>\tint$. 
\begin{eqnarray}
h_A(\bx,t)=\sum _{n=1}h_A^{(n)}(\bx,t)~,
%\label{eq-hSerie}
\end{eqnarray}
where
\begin{eqnarray}
h_A^{(1)}(\bx,t)&=&\int_{\Aint} dt' d\bx' \mG(\bx,t;\bx',t')S(\bx',t')\sim \frac{1}{(\gamma v)}~, \\
h_A^{(n)}(\bx,t)
&=&\int_{\Aint} dt' d\bx'\mG(\bx,t;\bx',t') \Delta W(\bx',t')h_A^{(n-1)}\sim \frac{1}{(\gamma v)^n}\qquad n\ge 2~.
\end{eqnarray}
This is because the limit of integration ignores the dashed orange box in Fig. \ref{fig-Int}(b), which will be given by $h_B$. The convenient way to describe $h_B$ is in the rest frame of the leaving solitary wave $\phi_B$. The calculation for that is identical to the above, just switching the role of the two solitary waves. That is what we meant by our earlier statement that the full result of the collision is to be described by the combination of two such calculations, one in the rest frame of each solitary wave. 

Let us be a bit more precise about the technical combination of $h_A$ and $h_B$ using the idea of initial data. We begin by taking the time slice at 
\begin{equation}
t=\tint = v^{-1}(L_A+\gamma^{-1}L_B)~,
\end{equation}
and considering everything to the right hand side of the solitary wave $\phi_A$ including itself,
\begin{equation}
x>-L_A/2~.
\end{equation}
The initial data on this semi-infinite slice is conveniently given by $h_A(\bx,\tint)$. The remaining half is supplemented by similar initial conditions in the rest frame of $\phi_B$. If we can correctly calculate both, then combining the two provides sufficient initial data to determine the results in the future. The future evolution happens in the background where the two solitary waves are far apart. Hence we can extend the validity of equations (\ref{eq-hSource}) and (\ref{eq-generalRecur}) to $t>\tint$ for $h=h_A$ and $h=h_B$ separately.

Let us once again emphasize the physical meanings of $h_A$ and $h_B$. $h_A$ does {\it not} represent the all-time result in the half spacetime region $\phi_A$ belongs to. It describes the perturbations {\it originated} from the half spacetime region at the time of collision. At later times, some of these perturbations will propagate outside this region, and this half also receives perturbations propagating from the other half where $\phi_B$ is.

The perturbations localized on separate solitary waves of course can be added linearly, since they never actually overlap. The region between two solitary waves is in empty space, and the propagating waves from both solitary waves are of small amplitudes. So they also add up linearly. The only exception is when a propagating wave from one solitary wave catches up with the other. That is of little concern to us, since the interaction between a small amplitude incoming wave and a solitary wave is part of the linearized dynamics of one solitary wave---it is not part of the collision dynamics that we want to deal with here. \footnote{One can even further consider the reflection of a mode with high enough momentum such that it comes back and catches up with the original solitary wave it was emitted from. That again can be treated by the intrinsic dynamics of one solitary wave and is separate from the short term effects of the collision.}
\section{Nonlinear evolution of Perturbations}
\label{sec-nonlinear}
In the main text, we only dealt with linearized equations of motion for the perturbations. Here we include a discussion that does not assume linearized perturbations.
For simplicity, we will limit ourselves to evaluating the perturbations at $t\sim \tint$. We will find that when we drop the linearization assumption, we have to correct the expressions for $h^{(n\ge 3)}$ provided in \eqn{eq-generalRecur}. We recommend that the reader refer to Sec. 2 for definitions and background on what is discussed below.

For a solution of the form
\begin{eqnarray}
\phi(\bx,t)=\phi_A(\bx,t)+\phi_B(\bx,t)+h(\bx,t)~,
\end{eqnarray}
the full, nonlinear equation of motion for $h$ is
\begin{eqnarray}
\Box h &=& V'(\phi_A+\phi_B+h)-V'(\phi_A)-V'(\phi_B)~, \nonumber \\
&=& W_0 h +  S(\bx,t) + \Delta W(\bx,t) h + \sum_{n=2}\frac{1}{n!}\left(\frac{d^{n+1}V|_{\phi_A+\phi_B}}{d\phi^{n+1}}\right)h^n~.
\end{eqnarray}
Expanding the perturbations $h$ as a series 
\Beq
h(\bx,t)=\sum_{n=1}^{\infty}h^{(n)}(\bx,t)~,
\Eeq
the order by order solutions can be written as 
\Beq
h^{(1)}(\bx,t)
&=&\int_{\Aint} dt' d\bx' \mG(\bx,t;\bx',t')S(\bx',t')~, \\
h^{(2)}(\bx,t)
&=&\int_{\Aint'}  dt' d\bx' \mG(\bx,t;\bx',t')
V''(\phi_A+\phi_B)h^{(1)}~,\nonumber\\
h^{(3)}(\bx,t)
&=&\int_{\Aint'}  dt' d\bx' \mG(\bx,t;\bx',t') 
\left[V''(\phi_A+\phi_B)h^{(2)}+\frac{1}{2}V'''(\phi_A+\phi_B)\left(h^{(1)}\right)^2\right]~,\nonumber\\
&\vdots& \nonumber
\Eeq
where $\mG=(\Box -W_0)^{-1}$ and $\Aint\sim \Aint'$ (see discussion in Sec. 2). For $n\le 2$, the expressions for $h^{(n)}$ above agree with those in \eqn{eq-generalRecur}. However, for $n\ge 3$, additional terms appear in the nonlinear case compared to the linearized one. For example, note that the term $\left(h^{(1)}\right)^2$ in the expression of $h^{(3)}$ above, is absent in the corresponding expression for $h^{(3)}$ in \eqn{eq-generalRecur}. Similar terms appear at higher orders as well. Note that these extra terms do not spoil our $h^{(n)}\propto \Aint h^{(n-1)}$ scaling. While not relevant for $n\le 2$, the nonlinear terms are important  for $n\ge 3$ and can be included in the calculation. 
%%%%%%%%%%%%%%%%%%%%%%%%%%%%%%%%%%%%%%%%%%%%%%%%
\section{Energy Conservation}
\label{sec-energy}

In scattering theory, the {\it optical theorem} often simplifies calculations considerably. In a classical field theory, the optical theorem is a direct consequence of energy conservation. In this appendix we show that the optical theorem is already useful at the leading order in $(\gamma v)^{-1}$. It shows that $B_0^{(1)}=0$ independently from the explicitly evaluation given in Sec.\ref{sec-pert}.  The general expression of the theorem requires analysis of higher order in the perturbation theory, which we postpone for future work. 

It is straightforward to compute the total energy of a single stationary kink plus small perturbations.\\ \\
\begin{eqnarray}
& & E_{\rm total \ stationary} \\ \nonumber 
&=&
\int dx~\left\{\frac{1}{2}\left(\partial_t[\phi_K(x) + h(x,t)]\right)^2
+\frac{1}{2}\left(\partial_x[\phi_K(x) + h(x,t)]\right)^2
+V\left[\phi_K(x) + h(x,t)\right]\right\}~, \\ \nonumber 
&=& \int dx~\left\{ \frac{1}{2}\left[\phi_K'(x)\right]^2 + V[\phi_K(x)]\right\} \\ \nonumber
& &\qquad+ \int dx~ \frac{1}{2}
\left\{\sum_a [f_a'(x)g_a(t)]^2+[f_a(x)g_a'(t)]^2+V''[\phi_K(x)][f_a(x)g_a(t)]^2\right\}~, \\ \nonumber
&=& M + \sum_a \frac{1}{2}E_a(A_a^2+B_b^2)~.
\end{eqnarray}
The equation of motion (\ref{eq-xeom}) ensures that the term linear in $h$ vanishes. We used the mode expansion $h=\sum_a f_ag_a$, with $g_a(t)=A_a \cos(\sqrt{E_a}t)+B_a \sin(\sqrt{E_a}t)$ in absence of sources. We will keep using the notation that all the $'$ are usual derivatives acting one functions of single variable.

%Recall that $E_a$'s are positive semi-definite. In the rest frame of one stationary kink total energy is not conserved. Since before the collision we have $A_a=B_a=0$ but after they will be nonzero, total energy increases. We should not be alarmed by this. It is simply because we treated the incoming kink as a fixed background source whose dynamics is not affected by the collision. Taking this into account, the incoming kink will lose energy and total energy will still be conserved. 

For the collision, it is convenient to go to the center-of-mass frame where the  two colliding kinks are on equal footing. In the center-of-mass frame, we have
\begin{equation}
\phi(x,t<0) = \phi_K[\gamma_c(x-v_ct)] + \phi_K[\gamma_c(x+v_ct)]
\end{equation}
before the collision. The center-of-mass boost factor $\gamma_c$ and velocity $v_c$ are related to those of the incoming kink in the stationary frame by
\begin{equation}
\gamma = 2\gamma_c^2-1\approx2\gamma_c^2~.
\end{equation}
After collision, we have
\begin{eqnarray}
\phi(x,t>0) &=& \phi_K\left[\gamma_c(x-vt)\right] 
              + h[\gamma_c(x-v_ct),\gamma_c(t-v_cx)] \\ \nonumber 
					&+&	\phi_K\left[\gamma_c(x+vt)\right] 
              + h[\gamma_c(x+v_ct),\gamma_c(t+v_cx)]~, \\ \nonumber
&=& \phi_K\left[\gamma_c(x-v_ct)\right] + 
\sum_a f_a\left[\gamma_c(x-v_ct)\right] G_a\left[\gamma_c(t-v_cx)\right] \\ \nonumber
&+& \phi_K\left[\gamma_c(x+v_ct)\right] + 
\sum_a f_a\left[\gamma_c(x+v_ct)\right] G_a\left[\gamma_c(t+v_cx)\right]~.
\end{eqnarray}
Long after the collision, the two kinks are far apart. Following the prescription in Appendix \ref{sec-LongTime}, for all practical purposes we can treat them independently. The energy for each solitary wave includes the energy of localized perturbations and waves propagating away from the soliton. Symmetry also guarantees that they each carries half of the total energy. \footnote{Even for kink-antikink, the energy density is still symmetric.}
\begin{equation}
\frac{1}{2}E_{\rm total \ CoM} = \gamma_c M + E_{\rm 1st} + E_{\rm 2nd}~.
\end{equation}
Here $E_{\rm 1st}$ and $E_{\rm 2nd}$ are the corrections to the energy at the first and second order $h$ respectively. Before we move on to analyze them further, note that conservation of total energy means $E_{\rm 1st} + E_{\rm 2nd}=0$, since the total energy has always been $2\gamma_cM$. Recall that $h$ will be given by a power series of $(\gamma v)^{-1}$, conservation of energy provides a cross-order relation at every order. This is commonly known as the optical theorem in perturbation theory.

We will postpone the full scope of the optical theorem to future work. Here we will demonstrate its power at the leading order. Recall that in Sec. \ref{sec-pert} we explicitly evaluated $B_0^{(1)}=0$ which means no leading order velocity change during the collision. Conservation of energy can give us that answer without an explicit evaluation. The process is a bit technical but the logic is quite simple. $B_0^{(1)}$ is going to be the only term contributing at the leading order in the energy conservation equation, $E_{\rm 1st} + E_{\rm 2nd}=0$. Therefore it must be zero.

In order to show that we focus on $E_{\rm 1st}$, since $E_{\rm 2nd}$ is automatically a higher order term.
\begin{eqnarray}
E_{\rm 1st} &=& \int dx~\left(\gamma_c\phi_K'[\gamma_c(x+v_ct)]\frac{\partial h}{\partial x} +\gamma_c v_c\phi_K'[\gamma_c(x+v_ct)]\frac{\partial h}{\partial t} +\frac{dV}{d\phi}h\right)~. \label{eq-E1} 
\end{eqnarray}
Unlike a stationary kink, the equation of motion (\ref{eq-xeom}) does not make $E_{\rm 1st}$ zero. Before we figure out what it is, let us clarify a few technical details.

We focus on the left moving kink, for which 
\begin{equation}
h = h[\gamma_c(x+v_ct),\gamma_c(t+v_cx)] = 
\sum_a f_a\left[\gamma_c(x+v_ct)\right] G_a\left[\gamma_c(t+v_cx)\right]~.
\label{eq-hvar}
\end{equation}
From this point on we will omit all arguments to make equations shorter. Since all the functions have one argument only, it should be clear what has been omitted. Again $'$ always means the usual derivative with respect to the single nontrivial argument of the corresponding function. Below, we provide a complete list of how such notation is related to the $\partial_x$ and $\partial_t$.
\begin{eqnarray}
\partial_x\phi_K &=& \gamma_c\phi_K'~, 
\ \ \ \ \ \partial_t\phi_K = \gamma_cv_c\phi_K'~, \\
\partial_xf_a &=& \gamma_cf_a'~, 
\ \ \ \ \ \ \partial_tf_a = \gamma_cv_cf_a'~, \\
\partial_xG_a &=& \gamma_cv_cG_a'~, 
\ \ \ \partial_tG_a = \gamma_cG_a'~.
\end{eqnarray}

The equation of motion (\ref{eq-xeom}) does not make $E_{\rm 1st}$ manifestly zero, but it does provide a useful equation,
\begin{eqnarray}
\int dx~ V' f_a G_a = \int dx~\phi_K'' f_a G_a
=-\int dx~\phi_K'(f_a'G_a+v_cf_aG_a')~,
\label{eq-phi''}
\end{eqnarray}
which results in a more compact expression,
\begin{equation}
E_{\rm 1st}=\int dx~\sum_a
\left[ 2\gamma_c^2 v_c^2\phi_K'f_a'G_a + \gamma_c^2 v_c(1+v_c^2)\phi_K'f_aG_a'\right]~.
\label{eq-E1st}
\end{equation}
Although the integral is nonzero, the contribution from all non-zero modes is zero. To show that, we need to use the property of the modes (\ref{eq-timeInd}),
\begin{equation}
f_a\phi_K'G_a' = E_a^{-1}(-f_a''+V''f_a)\phi_K'G_a'~.
\label{eq-nonzero}
\end{equation}
Integrating both sides and noting that both terms on r.h.s. can be integrated by parts,
\begin{eqnarray}
\int dx~V''\phi_K'f_aG_a' &=&
 \int dx~\frac{1}{\gamma_c}\partial_x V'f_aG_a' ~,\nonumber\\
&=& \int dx~\left(-V'f_a'G_a'-v_cV'f_aG_a''\right)~, \nonumber \\ \nonumber
&=& \int dx~\left(-\phi_K''f_a'G_a'+v_cE_a\phi_K''f_aG_a\right) ~. \\
\end{eqnarray}
Similarly,
\begin{eqnarray}
-\int dx~f_a''\phi_K'G_a' &=& 
\int dx~\left(f_a'\phi_K''G_a'+v_cf_a'\phi_K'G_a''\right) \\ \nonumber
&=& \int dx~\left(f_a'\phi_K''G_a'-v_cE_af_a'\phi_K'G_a\right)~.
\end{eqnarray}
In the last step of both equations above, we used the fact that after the collision $G_a$ obeys equation (\ref{eq-ga}), again in line with our discussion in Appendix \ref{sec-LongTime}. By combining them and plugging then back into equation (\ref{eq-nonzero}), we get
\begin{equation}
\int dx~\phi_K''f_aG_a = \int dx~\phi_K'(v_c^{-1}f_aG_a'+f_a'G_a)~.
\label{eq-trick}
\end{equation}
Combining equation (\ref{eq-trick}) and (\ref{eq-phi''}), we see that $E_{\rm 1st}$ in equation (\ref{eq-E1st}) gets exactly zero contributions from the nonzero modes.

For the zero mode, we first use the definition $\phi_K'=\sqrt{M}f_0$, and then integrate by parts to get
\begin{equation}
\int dx~2f_0f_0'G_0 = \int dx~v_cf_0^2G_0'~.
\label{eq-B0ibp}
\end{equation}
Combining equation (\ref{eq-B0ibp}) and (\ref{eq-E1st}), we get
\begin{equation}
E_{\rm 1st} = -\gamma_c\sqrt{M}v_cB_0~.
\label{eq-E1stend}
\end{equation}
This proves that $B_0^{(1)}$ is the only term at the leading order of $E_{\rm 1st}$, consequently the only term at the leading order of the conservation of energy equation. Therefore it must be zero.

\end{document}